\documentclass[graybox]{svmult}

\usepackage{type1cm}        
\usepackage{makeidx}         
\usepackage{graphicx}        
\usepackage{multicol}        
\usepackage[bottom]{footmisc}

\usepackage{newtxtext}       %
\usepackage[varvw]{newtxmath}       

\makeindex             

\begin{document}

\title*{On Fractional Generalizations of the Logistic Map and their Applications}
\titlerunning{Fractional Logistic Map}  
\author{Mark Edelman\orcidID{0000-0002-5190-3651}}
\institute{Mark Edelman \at 
Department of Physics, Stern College at Yeshiva University \\
245 Lexington Ave, New York, NY 10016, USA \\
Courant Institute of Mathematical Sciences, New York University \\
251 Mercer St., New York, NY 10012, USA \\ 
\email{edelman@cims.nyu.edu}}
%
\maketitle

\abstract*{The regular logistic map was introduced in 1960s, served as an example of a complex system, and was used as an instrument to demonstrate and investigate the period doubling cascade of bifurcations scenario of transition to chaos. In this paper, we review various fractional generalizations of the logistic map and their applications.}

\abstract{The regular logistic map was introduced in 1960s, served as an example of a complex system, and was used as an instrument to demonstrate and investigate the period doubling cascade of bifurcations scenario of transition to chaos. In this paper, we review various fractional generalizations of the logistic map and their applications.}

\section{Introduction}
\label{sec:1}
In 2010 I was invited to participate in IV International Conference
"Frontiers of Nonlinear Physics" on a ship. The ship departed from Nizhny Novgorod and a week later arrived at St. Petersburg. On the first day on the Volga River, Prof. Luo introduced himself and told me about his great respect for George Zaslavsky. George was a great physicist and a great human being. From 1995 to his tragic death in November of 2008, I worked with him at Courant Institute, which for a scientist is the best in the world place to work. Prof. Luo knew about our work and co-organized with George a series of conferences. He invited me to participate in some of the conferences and journals. Since then, I have had many meetings with Albert Luo, and he became my and my wife's friend. His acceptance and support of my research became a source of great inspiration for me. Albert once told us that when he was a student, his dream was to become a great scientist. Looking back at his scientific achievements and contributions to the scientific community, I may say that his dream has come true. Happy birthday, Albert. 

In George Zaslavsky's last paper, co-authored with Vasily Tarasov \cite{ZT1}, the authors introduced the universal fractional map in the way similar to the way in which the regular universal map is introduced by integrating equations of motion of a periodically kicked system. George emphasized the importance of the research on fractional maps for investigation of general properties of fractional systems. In my first paper written after George's death, co-authored with Vasily Tarasov \cite{ME2}, we investigated the fractional standard map. We found some key properties of fractional maps, like cascade of bifurcations type attractors (CBTT), power-law convergence of trajectories to stable periodic points, etc. This became the foundation for the further research that I pursued during the following 15 years. The methods used in 2008 \cite{ZT1} allowed derivation of the fractional maps of the orders $\alpha>1$ only. Therefore, the first numerically investigated maps were the standard maps of the orders $1<\alpha<2$. The results of the investigation were published in papers \cite{ME2,TE1,ME3,ME4}. The ways to derive fractional maps were described by Vasily Tarasov in journal papers \cite{T1,T2} and reviewed in Chapter 18 of \cite{TB}.

The situation changed in 2013 after a way to derive fractional maps of any non-negative orders was introduced in \cite{ME5,ME5n}. Then, it became possible to introduce the fractional logistic map as the solution of a differential equation describing a kicked system. The fractional logistic and the standard maps of the orders $0 \le \alpha \le 3$ were investigated in papers \cite{ME5,ME5n} and reviewed in \cite{ME6}. 

The next significant step in the development and investigation of the fractional logistic map was due to the introduction of fractional difference maps as solutions of fractional difference equations using Lemma 2.4 from \cite{CLZ}. 

Solutions of equations of kicked systems and fractional difference equations are not the only ways to introduce fractional maps. The other ways that we should mention are numerical schemes to solve differential equations and writing regular maps as fractions with the denominator, like $1+a^2$, which converges to a regular map in the case $a=0$. In this paper, the author will try, to the best of his knowledge, to review the existing versions of fractional logistic maps and their applications.

\section{The Regular Logistic Map}
\label{sec:2}

The first mentioning of the map that later was named the logistic map which the author was able to find is Eq.~(3) from \cite{Lorenz}. In this paper, the map was introduced by Edward Lorenz as the governing equation "capable of generating a stable climate". In 1976, Sir Robert M. May published in Nature a paper "Simple mathematical models with very complicated dynamics" \cite{LM5}, which became one of the
most cited papers - 4608 citations are registered by the Web of Science 
at the moment I am writing this sentence. In this review, the author, using the logistic map as an example, described the universal behavior typical for all nonlinear systems: transition to chaos through the period-doubling cascade of bifurcations. The main applications considered by the author were biological (even the variable used in the text was treated as "the population"), economic, and social sciences. 
This map has been used as a playground for investigations of one of the essential properties of nonlinear systems - the transition from order to chaos through a sequence of period-doubling bifurcations, which is called the cascade of bifurcations, and scaling properties of the corresponding systems (see
\cite{LM1,LM2,LM3,LM4,LM6}). 

\begin{figure}[!t]
\includegraphics[width=1.0\textwidth]{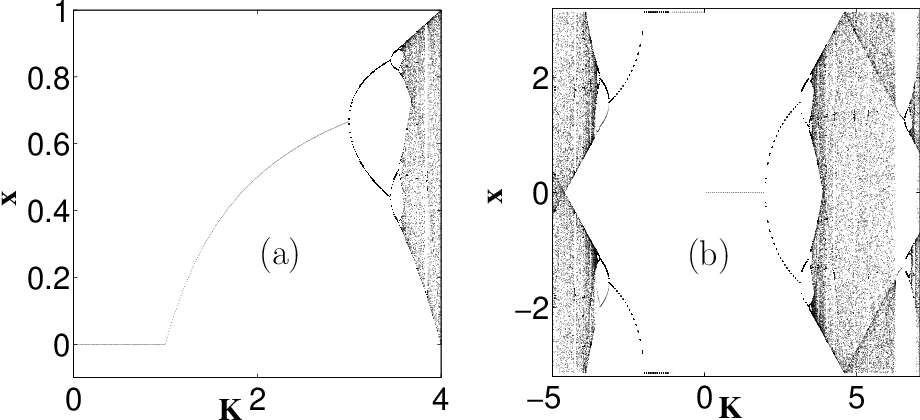}
\vspace{-0.25cm}
\caption{(a) The bifurcation diagram for the logistic map 
$x_{n+1}=Kx_n(1-x_n)$. 
(b) The bifurcation diagram for the 1D standard map (the circle map with the zero driving phase) $x_{n+1}= x_n - K \sin (x_n)$, $({\rm mod} \ 2\pi )$. 
This figure is reprinted from \cite{ME6} with the permission of Springer
Nature.
}
\label{BD1D}
\end{figure}

The stability properties of the logistic map (see \cite{LM5}) 
\begin{equation}
x_{n+1}=K x_{n}(1-x_{n})
\label{LM}
\end{equation}
for $0<K<4$ are summarized in the bifurcation diagram in Fig.~\ref{BD1D}(a). 
The $x=0$ fixed point (sink) is stable for  $K < 1$, 
the $(K-1)/K$ fixed point (sink) is stable for $1 < K < 3$, the  $T=2$ 
sink is stable for $3 \le K < 1-\sqrt{6} \approx 3.449$, the $T=4$ sink is stable when $3.449 < K < 3.544$, and the onset of chaos as a result of the period-doubling cascade of bifurcations occurs at $K  \approx 3.56995$. The bifurcation diagram for the logistic map is typical for almost all maps, see, e.g., the bifurcation diagram Fig.~\ref{BD1D}(b)  for the one-dimensional standard map \cite{ME5}.

\subsection{One-dimensional generalizations}
\label{1DStLog}

Since the introduction of the logistic map, various versions of generalized integer order logistic maps (and their fractional generalizations) were proposed by various authors. 

The simplest generalization would be \cite{FrRR}
\begin{equation}
x_{n+1}=K x^p_{n}(1-x^q_{n}), \ \ \ x \in [0,1], \ \ \ p,q>0.
\label{LMRR}
\end{equation}
The map's simplicity was the major motivation for its introduction. 

In \cite{FrCry}, the authors proposed the following generalization:
\begin{equation}
x_{n+1}=
\begin{cases}
&\frac{-q}{p^2}(p-x_n)^2+q, \ \ \ \ 0 \le x_n \le p,
\\
&\frac{-q}{(1-p)^2}(p-x_n)^2+q, \ \ \ \ p < x_n \le 1,
\end{cases}
\label{FrCry}
\end{equation}
to be used in cryptography.

The generalization proposed in \cite{FrFr1} of the generalized complex logistic map
\begin{equation}
z_{n+1}= z_n+\frac{\mu^\alpha}{\Gamma(1+\alpha)}(\beta   r z_{n}(a-bz_{n})+(1-\beta)z_n)
\label{frfr}
\end{equation}
to generate fractals, was fractionally generalized in \cite{FrFr2}.

There were more generalizations, but we will mention here only one other generalization which was investigated, along with its fractional counterpart in \cite{FrB1}. i.e.
\begin{equation}
x_{n+1}=\frac{K x_{n}(1-x_{n})}{1+r K x_{n}(1-x_{n})}.
\label{LMB1}
\end{equation}
This map does not blow up and iterates are bound over the entire real line unless one starts at the pole where the denominator is equal to zero.

\subsection{Two- and three-dimensional logistic maps}
\label{2DStLog}

All 2D logistic maps introduced by various researchers that the author was able to find are combinations of two one-dimensional logistic maps with various kinds of couplings. 

The situation is different when we define fractional and fractional difference $\alpha$-families of maps (see papers \cite{ME5,ME5n,ME7,ME8} and reviews \cite{ME6,HBV2,HBV4}). Natural fractional extensions of regular maps may be defined for any fractional order and properties of fractional maps are related to the properties of the corresponding integer order maps. It is interesting that the two-dimensional logistic map 
\begin{eqnarray}
\begin{cases}
& p_{n+1}= p_n+Kx_n(1-x_n)-x_n, 
\\
& x_{n+1}= x_n + p_{n+1}
\end{cases}
\label{LFMalp2}
\end{eqnarray}
is a quadratic area-preserving map. Its phase space contains stable elliptic islands and chaotic areas (no attractors).  
Quadratic area preserving maps which have a stable fixed point at the origin were investigated by H\'enon \cite{Henon69}  (for a recent review
on 2D quadratic maps see \cite{ZeraS2010}, for the general properties of 2D quadratic systems, see Albert Luo's book \cite{Albert}). The investigation the 2D member of the logistic $\alpha$-family of maps, is based on the analysis of the evolution of its periodic
points with the increase of the map parameter $K$.  
For $K \in (-3,1)$, the map Eq.~\eqref{LFMalp2} has the stable fixed point
$(0,0)$  which turns into the  fixed point $((K-1)/K,0)$ stable 
for  $K \in (1,5)$.
The $T=2$ elliptic point
{\setlength\arraycolsep{0.5pt}
\begin{eqnarray}
\begin{cases}
&x = \frac{K+3 \pm \sqrt{(K+3)(K-5)}}{2K},  
\\ 
&p=\pm \frac{\sqrt{(K+3)(K-5)}}{K}
\end{cases}
\label{LFMalp2T2}
\end{eqnarray} 
}
\begin{figure}[!t]
\includegraphics[width=1.0\textwidth]{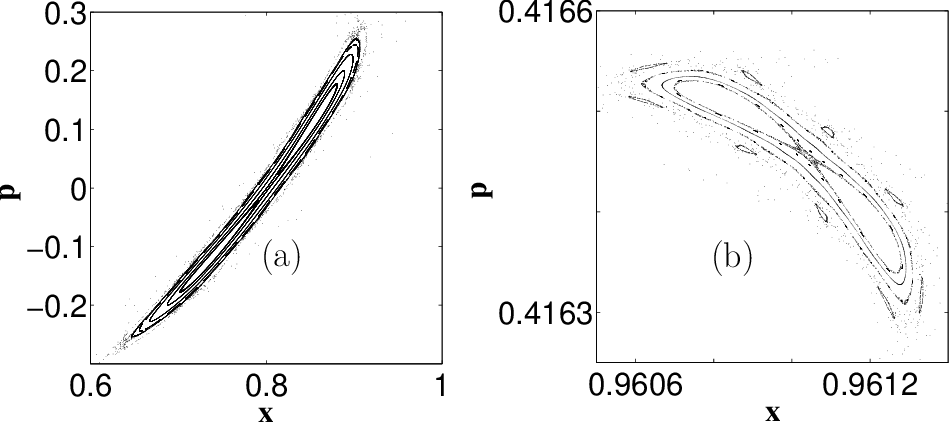}
\vspace{-0.25cm}
\caption{ 
Bifurcations in the 2D Logistic Map:
(a) $T=1$ $\rightarrow$ $T=2$ bifurcation at $K = 5$ ($K=5.05$ 
on the figure). 
(b) $T=8$ $\rightarrow$ $T=16$ bifurcation at $K \approx 5.5319$  ($K=5.53194$ 
on the figure). This figure is reprinted from \cite{ME6} with the permission from Springer Nature.  
}
\label{FigLog2D}
\end{figure}
is stable for $-2 \sqrt{5}+1<K<-3$ and $5<K<2 \sqrt{5}+1$.
The period doubling cascade of bifurcations (for $K>0$) 
follows the scenario of the elliptic-hyperbolic point transitions with the births of the double periodicity islands inside the original island 
which has been investigated in \cite{Schmidt} and applied to an investigation of the standard map's stochasticity at low values of the map's parameter.
Further bifurcations in the 2D logistic map, 
$T=2$ $\rightarrow$ $T=4$ at $K \approx 5.472$,
$T=4$ $\rightarrow$ $T=8$ at $K \approx 5.527$, $T=8$ $\rightarrow$ $T=16$
at $K \approx 5.5319$, $T=16$ $\rightarrow$ $T=32$ at $K \approx 5.53253$,
etc., and the corresponding decreases of the areas of the islands of stability (see Fig.~\ref{FigLog2D}) lead to chaos.

The universal 3D fractional map is defined as 
\begin{eqnarray}
\begin{cases}
&x_{n+1}=  x_n+y_{n+1}-\frac{1}{2}z_{n+1},   
\\
&y_{n+1}=y_n+z_{n+1}, \label{3DUMn} \\
&z_{n+1}=-G_K(x_n)+z_n, 
\end{cases}
\end{eqnarray} 
which is a volume preserving map. If the generalized map is 
\begin{equation}
x_{n+1}=f(x),
\label{Gen1D}
\end{equation}
then, in order for the universal fractional map in the 1D case to converge to this map, $G_K(x)$ must be defined as (see, e.g. \cite{HBV2}) 
\begin{equation}
G_K(x)=x-f(x).
\label{Gen1D_G}
\end{equation}

This map has fixed points $z_0=y_0=G_K(x_0)=0$ and
stability of these points can be analyzed by considering the eigenvalues 
$\lambda$ of the
matrix (corresponding to the tangent map)
\begin{equation}
\left( \begin{array}{ccc}
1-0.5\dot{G_K}(x_0) & 1 & 0.5 \\
-\dot{G_K}(x_0) & 1 & 1 \\
-\dot{G_K}(x_0) & 0 & 1 \end{array} \right).
\label{3DM}
\end{equation}
The only case in which the fixed points could be stable is  $\dot{G_K}(x_0)=0$,
when $\lambda_1=\lambda_2=\lambda_3=1$. From Eq.~(\ref{3DUMn}), it follows that 
the only $T=2$ points are the fixed points.

Eq.~(\ref{3DUMn}) with $G_K(x)=x-Kx(1-x)$ produces
the 3D logistic map
\begin{eqnarray}
\begin{cases}
&x_{n+1}=  x_n+y_{n+1}-\frac{1}{2}z_{n+1},  
\\
&y_{n+1}=y_n+z_{n+1}, \label{3DLMn} \\
&z_{n+1}=Kx_n(1-x_n)-x_n+z_n. 
\end{cases}
\end{eqnarray} 
Three-dimensional quadratic volume preserving maps were investigated in 
\cite{Moser1994,LoMeiss1998}.

\section{Defining the fractional logistic map}
\label{sec:3}

There are many various ways to define a fractional map. The most popular way is to define a map as a solution of a fractional difference equation. The simplest equation that may be used to define a fractional order $\alpha$ map is 
\begin{equation}
\Delta^\alpha_{a,h} x=f(x),
\label{FDE1}
\end{equation}
where $\Delta^\alpha_{a,h}$ is a generalization of the integer ($\alpha=1$)  forward $h$-difference operator   
\begin{equation}
\Delta_h f(x)=f(x+h)-f(x).
\label{DF1}
\end{equation}
$h$ is frequently assumed to be 1. Because fractional operators are integro-differential or summation-difference operators, $a$ stands for the initial point of summation. In many problems, $a=0$ and $h=1$, in which case, we will omit the subscripts. Even though the maps are called fractional, the corresponding operators are defined for any real ($\alpha \in \mathbb{R}$) or complex ($\alpha \in \mathbb{C}$) values of $\alpha$.
Some authors prefer generalizations of the backward difference operator $\nabla_{h} f(x)=f(x)-f(x-h)$. There is an obvious relationship between fractional $\Delta$ and $\nabla$ operators (see \cite{AE2009N} and     Eqs.~(46)~and~(47) from \cite{HBV2}):
\begin{eqnarray}
&& \Delta^{\alpha}_ay(t-\alpha)= \nabla^{\alpha}_{a-1}y(t); \  \ t\in \mathbb{N}_{m+a}, \\
&&  \Delta^{-\alpha}_ay(t+\alpha)= \nabla^{-\alpha}_{a-1}y(t); \  \ t\in \mathbb{N}_{a},
\label{DeltaNabla} 
\end{eqnarray} 
where $\mathbb{N}_{a}=\{a,a+1, a+2, ...\}$. As a result, solutions of the fractional difference $\Delta$ (Caputo-like, see \cite{Anas}) and $\nabla$ equations (Lemma 2.4 from \cite{DifSum}),
\begin{equation}
^C\Delta^{\alpha} x(n) = -G_K(n+\alpha-1,x(n+\alpha-1))
\label{LemmaDif}
\end{equation}  
and (see \cite{Nabla1})
\begin{equation}
\nabla^{\alpha} x(n+1) = -G_K(n,x(n)),
\label{LemmaDif1}
\end{equation}  
where $K$ is a parameter, $0<\alpha<1$, and $n\in \mathbb{N}_{0}$, with the initial condition
\begin{equation} 
x(0) =   x_0, 
\label{ICFalFacMap}
\end{equation}
are identical, and they may be written as a fractional difference map with falling factorial-law memory (Eq.~(59) from \cite{HBV2})
\begin{equation} 
x_{n} =   x_0 -\frac{1}{\Gamma(\alpha)}  
\sum^{n-1}_{s=0}(n-s-2+\alpha)^{(\alpha-1)} 
G_K(s,x_{s}),
\label{FalFacMap}
\end{equation}
or in the equivalent form (Eq.~(4) from \cite{Nabla2})
\begin{equation} 
x_{n} =   x_0 - 
\sum^{n-1}_{s=0}B(n-1,\alpha;s)G_K(s,x(s)),
\label{FalFacMapEqv}
\end{equation}
where 
\begin{equation}
B(n,\alpha;s)=
\left( 
\begin{array}{c}
n-s+\alpha-1 \\ n-s
\end{array} \right) 
\label{Bdef}
\end{equation} 
for $0\le s \le n$.
The falling factorial function is defined as 
\begin{equation}
t^{(\alpha)} =\frac{\Gamma(t+1)}{\Gamma(t+1-\alpha)}, \ \ t\ne -1, -2, -3,
....
\label{FrFac}
\end{equation}
Falling factorial-law memory is asymptotically power-law memory, i.e.
\begin{equation}
\lim_{t \rightarrow
  \infty}\frac{\Gamma(t+1)}{\Gamma(t+1-\alpha)t^{\alpha}}=1,  
\ \ \ \alpha \in  \mathbb{R},
\label{GammaLimit}
\end{equation}
and $\Gamma(\alpha)$ is the gamma function. 

Considering the identity of maps arising from the use of fractional forward and backward difference ($\Delta$ and $\nabla$) operators, in what follows, we will utilize only the fractional forward difference operator $\Delta$.

\subsection{First fractional logistic maps}
\label{sec:3.1}

To the best of our knowledge, the first definition of a fractional difference operator was given by Diaz and Osier \cite{DO}
as a natural extension of the regular difference. The value of the difference in this definition
\begin{equation}
\Delta^{\alpha}f(z)= \sum^{\infty}_{k=0}(-1)^k
\left( 
\begin{array}{c}
\alpha \\ k
\end{array} \right) 
f(z+\alpha-k)
\label{DOdef}
\end{equation} 
depends on past values of function starting from the values at negative infinity. This is not practical. The next, more practical, definition using Taylor's series was given in Hirota's publication \cite{Hirota}. The latter definition was modified by Nagai in papers \cite{Nagai1,Nagai2}. In the ArXiv version of this paper \cite{Nagai1}, titled ``Fractional logistic map'', the author presented the first version of the fractional logistic map, which was his fractional generalization of the map
\begin{equation}
\frac{u_{n+1}-u_n}{\varepsilon}=a u_{n}(1-u_{n})  \   \ (a>0).
\label{LM1}
\end{equation}
Later, the author realized that this map is not the logistic map but rather a discrete form of the logistic equation, and the journal version of the paper \cite{Nagai2} did not mention the logistic map. We should note here that confusing the logistic map with the discrete logistic equation is a relatively common thing. For example, the first published paper on the fractional difference logistic map (FDLM) \cite{FallLog} has nothing to do with the logistic map -- it introduces the discrete version of the fractional logistic difference equation. Unlike Nagai, the authors of \cite{FallLog} continue citing their map as the fractional logistic map (FLM) even after the comments on this paper outlining the authors mistakes were published in the same journal Nonlinear Dynamics \cite{Yuxi}. 
The form of the map proposed by Nagai allowed him to integrate the mapping analytically. 

The next FLM was proposed by Stanislavsky \cite{Stan}. He directly introduced power-law memory into the equation of the logistic map using a robust algorithm of numerical fractional integration \cite{Sref}. The thing correctly noticed by Stanislavsky is the dependance of fractional bifurcation diagrams on the fractional order $\alpha$ of the map. But Stanislavsky's map turns into the regular logistic map when the order $\alpha$ is zero (not one). Another mistake made by the author was not removing the results of initial iterations (transient processes) from the set of data used to draw the diagrams. Fractional maps converge to the asymptotic fixed points according to the power law with low powers (the power is $-\alpha$ in the case of fractional or fractional difference maps). As a result, bifurcation diagrams strongly depend on the number of iterations and the initial conditions. Transient processes imbedded into bifurcation diagrams of fractional maps make them messy and unusable (the same mistake made in \cite{FallLog} was analyzed in \cite{Yuxi}).   

In 2013, Munkhammar \cite{Mun} defined a fractional logistic map as 
\begin{equation}
x_{n+1}=\frac{\lambda}{\Gamma(\alpha+2)} \Bigl(1-\frac{2x_n}{\alpha+2}\Bigr) x_{n}^{1+\alpha} 
\label{LMnm}
\end{equation}
simply because the right side of this map is an order $\alpha$ fractional integral of the right side of the regular logistic map. The resulting map Eq.~(\ref{LMnm}) has no memory and is not in the spirit of fractional calculus.

From the author's point of view, the most natural ways to introduce fractional generalizations of any map, are: a) to use solutions of 
differential equations of kicked systems, like the way in which the universal map is introduced in regular dynamics, and we will call such maps fractional maps; or b) to use solutions of fractional difference equations with the most accepted form of the fractional difference operator, and we will call such maps fractional difference maps. The first FLM, introduced in 2013 \cite{ME5}, will be discussed in the next section. After that, we will discuss the FDLM, which was introduced later.

\subsection{The fractional logistic map}
\label{sec:3.2}

The FLM is a particular form of the fractional universal map introduced in \cite{ZT1,ME5,ME5n}. As in the regular ($\alpha=2$) case, in the fractional case, the Caputo fractional universal map is a solution of the following equation of a periodically kicked system: 
\begin{equation}
_0^CD^{\alpha}_tx(t) +G_K(x(t- \Delta )) \sum^{\infty}_{n=-\infty} \delta \Bigl(\frac{t}{h}-(n+\varepsilon)
\Bigr)=0,    
\label{UM1D2DdifC}
\end{equation}
 where h is a period of the kicks, $\varepsilon > \Delta > 0$, $\varepsilon  \rightarrow 0$, $0 \le N-1
 < \alpha \le N$,  $\alpha \in \mathbb{R}$, $N \in \mathbb{Z}$, 
and the initial conditions 
\begin{equation}
(D^{k}_tx)(0+)=b_k,  \    \ k=0,...,N-1.
\label{UM1D2DdifCic}
\end{equation}
The left-sided Caputo fractional derivative $_0^CD^{\alpha}_t x(t)$ is defined for
$t>0$ \cite{SKM} as
\begin{equation}
_0^CD^{\alpha}_t x(t)=
\frac{1}{\Gamma(m-\alpha)}  \int^{t}_0 
\frac{ D^m_{\tau}x(\tau) d \tau}{(t-\tau)^{\alpha-m+1}},
\label{Cap}
\end{equation}
where $m-1 <\alpha \le m$.

Integration of Eq.~(\ref{UM1D2DdifC}) with the initial conditions Eq.~(\ref{UM1D2DdifCic}) yields the Caputo universal $\alpha$-family of maps ($\alpha$FM) (see \cite{TarBook,MEBr})
{\setlength\arraycolsep{0.5pt}
\begin{equation}
x_{n+1}= \sum^{N-1}_{k=0}\frac{b_k}{k!}h^k(n+1)^{k} 
-\frac{h^{\alpha}}{\Gamma(\alpha)}\sum^{n}_{k=0} G_K(x_k) (n-k+1)^{\alpha-1}.
\label{FrCMapx}
\end{equation}
}

In this paper, we will consider almost exclusively equations with the Caputo fractional derivatives and differences because they are the simplest and most accepted ones and have no problems with the definition of initial conditions typical for equations with the Riemann-Liouville derivatives and differences.

The FLM is obtained from Eq.~(\ref{FrCMapx}) when we make a substitution:
\begin{equation}
G_K(x) = x-Kx(1-x)
\label{FrLogGen}.
\end{equation}
Finaly, the Caputo logistic fractionsl $\alpha$-family of maps LF$\alpha$FM ($\alpha >0$) is written as
{\setlength\arraycolsep{0.5pt}
\begin{equation}
x_{n+1}= \sum^{N-1}_{k=0}\frac{b_k}{k!}h^k(n+1)^{k} 
-\frac{h^{\alpha}}{\Gamma(\alpha)}\sum^{n}_{k=0} [x_k-Kx_k(1-x_k)]
(n-k+1)^{\alpha-1}.
\label{FrCMapLog}
\end{equation}
}
The case relevant to most applications is $0< \alpha <1$. So, in what follows, we will consider mainly this case when
{\setlength\arraycolsep{0.5pt}
\begin{equation}
x_{n+1}= x_0 
-\frac{h^{\alpha}}{\Gamma(\alpha)}\sum^{n}_{k=0} [x_k-Kx_k(1-x_k)]
(n-k+1)^{\alpha-1}
\label{FrCMapLogLow}
\end{equation}
}
and call this map the Caputo FLM.

The investigated in \cite{ME5n,ME6} Caputo FLM with $1<\alpha<2$ and $h=1$ may be written as a 2D map 
\begin{eqnarray}
&&x_{n+1}=x_0+ p_0(n+1)^{k} 
-\frac{1}{\Gamma(\alpha)}\sum^{n}_{k=0} [x_k-Kx_k(1-x_k)] (n-k+1)^{\alpha-1}, 
\label{LMCx}  \\
&&p_{n+1}=p_0 
-\frac{1}{\Gamma(\alpha-1)}\sum^{n}_{k=0} [x_k-Kx_k(1-x_k)] (n-k+1)^{\alpha-2}.
\label{LMCp}
\end{eqnarray} 
   
\subsection{Fractional difference logistic map}
\label{sec:3.3}

As we mentioned at the beginning of Section~\ref{sec:3.1}, the first definitions of fractional difference operators were introduced in papers 
\cite{DO,Hirota,Nagai1,Nagai2}. In 1988 paper \cite{Gray}, Gray and Zhang introduced a fractional sum operator in the way, similar to the way in which the fractional integral was introduced by Liouville, by extending the Cauchy $n$-fold sum formula to all real values of variables. The proof of the Cauchy $n$-fold sum formula may be found, e.g., in Section 3 of \cite{ME8}. In 1989 paper \cite{Miller},  Miller and Ross introduced the discrete fractional sum operator based on the Green's function approach. The resulting definitions of the fractional forward sum 
operator
\begin{equation}
\Delta^{-\alpha}_{a}f(t)=\frac{1}{\Gamma(\alpha)} \sum^{t-\alpha}_{s=a}(t-s-1)^{(\alpha-1)} f(s)
\label{MRDef}
\end{equation}
obtained in \cite{Gray} and \cite{Miller} are identical.
As a result, this definition became commonly accepted and used by many researchers. 

The fractional (left) Caputo-like difference operator (see \cite{Anastas}) is defined as
{\setlength\arraycolsep{0.5pt}   
\begin{eqnarray} 
&&^C\Delta^{\alpha}_a x(t) =  \Delta^{-(m-\alpha)}_{a}\Delta^{m} x(t)
\nonumber \\
&&=\frac{1}{\Gamma(m-\alpha)} \sum^{t-(m-\alpha)}_{s=a}(t-s-1)^{(m-\alpha-1)} 
\Delta^m x(s).
\label{FDC}
\end{eqnarray}
}  
This definition is valid for any real $\alpha \ge 0$ (see \cite{HBV2}).

Fractional $h$-difference operators are generalizations of the fractional difference operators (see \cite{HBV2} and references [41-47] therein).
The h-sum operator is defined as
\begin{equation}
(_a\Delta^{-\alpha}_{h}f)(t)=\frac{h}{\Gamma(\alpha)} \sum^{\frac{t}{h}-\alpha}_{s=\frac{a}{h}}(t-(s+1)h)^{(\alpha-1)}_h f(sh),
\label{hSum}
\end{equation}
where $\alpha \ge 0$, $(_a\Delta^{0}_{h}f)(t)=f(t)$, $f$ is defined on
$(h\mathbb{N})_a$, and $_a\Delta^{-\alpha}_h$ on  
$(h\mathbb{N})_{a+\alpha h}$, $(h\mathbb{N})_t=\{t,t+h, t+2h, ...\}$.
The $h$-factorial $t^{(\alpha)}_h$ is defined as
\begin{equation}
t^{(\alpha)}_h =h^{\alpha}\frac{\Gamma(\frac{t}{h}+1)}{\Gamma(\frac{t}{h}+1-\alpha)}= h^{\alpha}\Bigl(\frac{t}{h}\Bigr)^{(\alpha)}, \ \ \frac{t}{h} \ne -1, -2, -3,
...
\label{hFrFac}
\end{equation}
With  $m=\lceil \alpha \rceil$   
the Caputo (left) h-difference is defined as
{\setlength\arraycolsep{0.5pt}
\begin{eqnarray}
&&(_a\Delta^{\alpha}_{h,*} x)(t) =  
(_a\Delta^{-(m-\alpha)}_h (\Delta^{m}_{h}x))(t) \nonumber \\  
&&=\frac{h}{\Gamma(m-\alpha)} \sum^{\frac{t}{h}-(m-\alpha)}_{s=\frac{a}{h}}(t-(s+1)h)^{(m-\alpha-1)}_h 
(\Delta^m_hx)(sh),
\label{hFDC}
\end{eqnarray}
}
where $(\Delta^{m}_{h}x))(t)$ is the $m$th power oh the forward $h$-difference operator 
\begin{equation} 
(\Delta_{h}x)(t)=\frac{x(t+h)-x(t)}{h}.
\label{FHD}
\end{equation} 
The following theorem was formulated in \cite{MEBr}:

\begin{theorem}
 For $\alpha \in \mathbb{R}$, $\alpha \ge 0$ the Caputo-like 
h-difference equation 
\begin{equation}
(_0\Delta^{\alpha}_{h,*} x)(t) = -G_K(x(t+(\alpha-1)h)),
\label{LemmaDif_n_h}
\end{equation}
where $t\in (h\mathbb{N})_{m}$, with the initial conditions 
 \begin{equation}
(_0\Delta^{k}_h x)(0) = c_k, \ \ \ k=0, 1, ..., m-1, \ \ \ 
m=\lceil \alpha \rceil
\label{LemmaDifICn_h}
\end{equation}
is equivalent to the map with $h$-factorial-law memory
\begin{eqnarray} 
&&x_{n+1} =   \sum^{m-1}_{k=0}\frac{c_k}{k!}((n+1)h)^{(k)}_h 
\nonumber \\
&&-\frac{h^{\alpha}}{\Gamma(\alpha)}  
\sum^{n+1-m}_{s=0}(n-s-m+\alpha)^{(\alpha-1)} 
G_K(x_{s+m-1}), 
\label{FalFacMap_h}
\end{eqnarray}
where $x_k=x(kh)$, 
which is called the $h$-difference Caputo universal  
$\alpha$-family of maps.
\label{T2}
\end{theorem}

With $G(x)$ defined by Eq.~(\ref{FrLogGen}), the Caputo fractional difference logistic map (FDLM) for $0<\alpha<1$ may be written as (see \cite{ME8,HBV4})
\begin{equation} 
x_{n+1} =  x_0  
\label{LMlt1} \\
-\frac{h^{\alpha}}{\Gamma(\alpha)}
\sum^{n}_{s=0}\frac{\Gamma(n-s+\alpha)}{\Gamma(n-s+1)}[x_s-Kx_s(1-x_s)]. 
\end{equation}

The $1<\alpha<2$ Caputo FDLM is
\begin{equation} 
x_{n+1} =  x_0 +h\Delta_h x(0) (n+1) -\frac{h^{\alpha}}{\Gamma(\alpha)}
 \label{LMgt1}  
\times
\sum^{n-1}_{s=0}\frac{\Gamma(n-s+\alpha-1)}{\Gamma(n-s)}[x_{s+1}-Kx_{s+1}(1-
x_{s+1})].
\end{equation}
After introduction $p_n=\Delta x_{n-1}$, the 
FDLM may be written as a 2D map
{\setlength\arraycolsep{0.5pt}   
\begin{eqnarray} 
&&p_{n} =  p_1 -\frac{K}{\Gamma(\alpha-1)}
\times \sum^{n}_{s=2}\frac{\Gamma(n-s+\alpha-1)}
{\Gamma(n-s+1)}[x_{s-1}-Kx_{s-1}(1-x_{s-1})], 
\label{LMgt1p}  \\
&& x_n=x_{n-1}+p_n,  \ \ n \ge 1.
\label{LMgt1x} 
\end{eqnarray}
}

\subsection{Generalized fractional logistic map}
\label{sec:3.4}

As we already mentioned, the FLM was introduced in 2013 \cite{ME5}. The FDLM was introduced later \cite{ME8}; but it became clear from the time of its introduction, that both maps may be written and investigated in a unified generalized form (see Secs.~2.3.2~and~2.3.3 from \cite{HBV4}, \cite{Cycles}, Eq.~(4.31) from \cite{Avigail}, and Sec.~3.1 from \cite{IFAC}) called the generalized FLM (GFLM): 
 \begin{equation}
x_{n+1}= g(n,h)
-h^{m-1}\sum^{n}_{k=0} G^{m-1}(x_k) U_{\alpha}(n-k+1),
\label{FrCMapxm}
\end{equation}
where $m=\lceil \alpha \rceil$, $\alpha \in \mathbb{R}$, $\alpha>0$, $h>0$, 
$\Delta^{m-1}U_{\alpha}(i) \in \mathbb{D}^{0}(\mathbb{N}_1)$, 
$i \in \mathbb{N}_{2-m}$, and
\begin{equation}
G^{m-1}(x)= [x-Kx(1-x)]h^{\alpha-m+1}/\Gamma(\alpha).
\label{G}
\end{equation}
For $0<\alpha<1$, $g(n,h)=x(0)=x_0$. For $1<\alpha<2$ 
\begin{equation}
g(n,h)=x_0+h (n+1) p_0+ h g_1(n),
\label{g}
\end{equation}
where $p_0$ is equal to $b_1$ defined by Eq.~(\ref{UM1D2DdifCic}) in the case of the  FLM or to $c_1$ defined by Eq.~(\ref{LemmaDifICn_h}) in the case of the FDLM, and $g_1(n)=0$ in the FLM or $g_1(n)=h^{\alpha-1}[x_0-Kx_0(1-x_0)](n-1+\alpha)^{(\alpha-1)}/\Gamma(\alpha)$ in the FDLM. 

The space 
$\mathbb{D}^i(\mathbb{N}_1)$, $i=0,1,2, ...$,
is defined in \cite{Avigail} as
{\setlength\arraycolsep{0.5pt}
\begin{eqnarray}
&&\mathbb{D}^i(\mathbb{N}_1) \ \ = \ \ \{f: \left|\sum^{\infty}_{k=1}\Delta^if(k)\right|>N, 
\nonumber \\
&& \forall N, \ \ N \in \mathbb{N}, 
\sum^{\infty}_{k=1}\left|\Delta^{i+1}f(k)\right|=C, \ \ C \in \mathbb{R}_+\}.
\label{DefForm}
\end{eqnarray}
}
$U_\alpha(n)=0$ for $n<1$. For $n>0$,   
\begin{equation}
U_\alpha(n)=n^{\alpha-1}
\label{Ufr}
\end{equation}
in fractional maps or
\begin{equation}
U_\alpha(n)=(n+\alpha-2)^{(\alpha-1)}
\label{Ufrdif}
\end{equation}
in fractional difference maps.

\section{Properties of the Fractional and Fractional Difference Logistic Maps}
\label{sec:4}

It is known that continuous and discrete fractional maps do not have periodic points except the fixed points \cite{PerD1,PerD2,PerC1,PerC2, PerC3,PerC4,PerC5}. For fractional difference maps, this was explicitly demonstrated in \cite{PerD3} (Theorem 5.1 in that paper). The fractional and fractional difference logistic maps have two obvious fixed points, $x=0$ and $x=(K-1)/K$. Initial investigation of fractional maps was concentrated on finding fixed and period two points, numerical analysis of types of behavior of single trajectories, and approximations of the asymptotic bifurcation diagrams based on numerical simulations of single trajectories.

\subsection{Asymptotically periodic and bifurcation points}
\label{sec:4.1}

The fixed points of the FLM and FDLM, $x=0$ and $x=(K-1)/K$, are obtained from the equation 
\begin{equation}
 G_K(x) = x-Kx(1-x)=0.
\label{FP}
\end{equation}

The algebraic equation defining asymptotically period two points in the FLM of the orders $1<\alpha<2$ for $h=1$ was derived in \cite{ME5}, i.e.
{\setlength\arraycolsep{0.5pt}
\begin{eqnarray}
&&x_{le}^2-\Bigl(\frac{2\Gamma(\alpha)}{KV_{\alpha l}}+
\frac{K-1}{K}\Bigr)x_{le}+  \Bigl(\frac{\Gamma(\alpha)}{2KV_{\alpha l}}+
\frac{K-1}{4K}\Bigr)^2   \nonumber \\ 
&&\hspace{-0.2cm} -\frac{(K-1)\Gamma(\alpha)}{K^2V_{\alpha l}}-
\frac{(K-1)^2}{2K^2}=0,
\label{x}
\end{eqnarray}
}
which for positive values of $K$ has solutions only when 
\begin{equation} \label{KcL} 
K  \ge K_{c1l}=1+\frac{2\Gamma(\alpha)}{V_{\alpha l}}.
\end{equation}
$K_{c1l}$ is a fixed point -- period two bifurcation point. The sum $V_{\alpha  l}$ was defined in \cite{ME5}, but in the following we will use defined in \cite{Cycles,Avigail,Bif} sums $S_{i,j}$, i.e.
{\setlength\arraycolsep{0.5pt}   
\begin{eqnarray} 
&&S_{j+1,l}={S}^{0}_{j+1,l}=\sum^{\infty}_{k=0}\Bigl[
U_{\alpha} (lk+j) - U_{\alpha} (lk+j+1)\Bigr], \   \ 0 \le j<l,
\label{Ser}
\end{eqnarray}
}
and $V_{\alpha  l}=2S_{2,2}$ ($U_\alpha()$ is defined by Eq.~(\ref{Ufr})).  

The following theorem was proven in \cite{Avigail} 
\begin{theorem}\label{Th1} 
In fractional maps of the orders $1<\alpha<2$, the total of all physical momenta of asymptotically period--$l$ points is zero:
\begin{equation}
\sum^{l}_{j=1}p_{lim,j}=0, \  \ l \in \mathbb{Z}, \  \ l>1.
\label{ClosePfn}
\end{equation}
\end{theorem}
This implies that the momentum of the asymptotically $T=2$ point in the FLM Eqs.~(\ref{LMCx})~and~(\ref{LMCp}) satisfies the following condition 
\begin{equation}
p_1=-p_2,
\label{PT2f}
\end{equation}
\begin{figure}[b]
\sidecaption
\includegraphics[scale=.5]{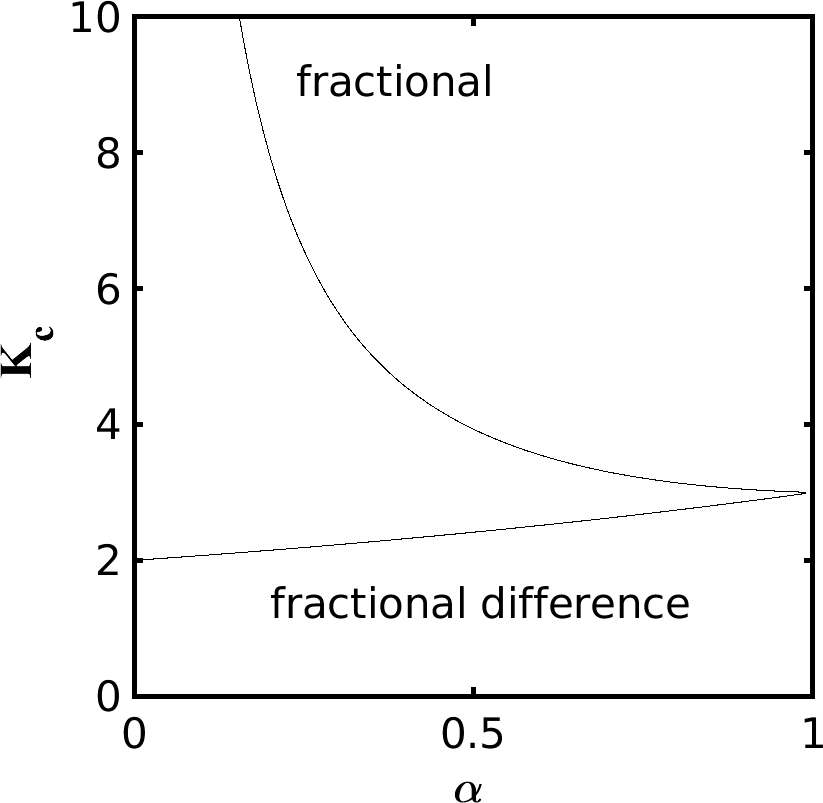}
\caption{Asymptotic bifurcation K-$\alpha$ curves (Eqs.~(\ref{KcL})~and~(\ref{Kcal})) on which transition from a fixed point to a $T=2$ cycle occurs for fractional (upper curve) and fractional difference logistic maps ($h=1$). This figure is reprinted from \cite{Cycles} with the permission of Springer Nature.} 
\label{fig3}       
\end{figure}
which is also valid for the FDLM.
A system of equations that defines the asymptotically $T=2$ point when $0<\alpha<2$ in both, the FLM and the FDLM, was derived in \cite{HBV4} (Eq.~(83) in that paper).
The solution of this system is 
\begin{equation}
x_{1,2}=\frac{K_{C1s}+K-1 \pm \sqrt{(K-1)^2-K_{C1s}^2}}{2K}
\label{eqT2logSolu},
\end{equation}
where 
\begin{equation} \label{Kcs} 
K_{C1s}=\frac{\Gamma(\alpha)}{h^{\alpha}S_{2,2}}.
\end{equation}
The asymptotically fixed point -- $T=2$ bifurcation points are defined by
\begin{equation} \label{Kcal} 
K_{c1l}=1 \pm \frac{\Gamma(\alpha)}{h^{\alpha}S_{2,2}}.
\end{equation}

\subsubsection{Asymptotically periodic points in the GFLM of the orders $\alpha>0$}
\label{sec:4.1.1}

Equations that define $l$ asymptotically $l$-periodic points (cycles of the period $l$) $x_{l,k}$, i.e.
\begin{equation}
x_{l,k}=\lim_{N \rightarrow
  \infty} x_{Nl+k}, \  \  \ 0<k<l+1,
\label{TlpointEx}
\end{equation}
in the GFLM, which include the FLM and the FDLM, of any positive order $\alpha>0$ were derived in \cite{Cycles,Avigail}, i.e.
$$ 
x_{lim,k+1}+(-1)^{m}x_{lim,k-m+1}-
\sum^{m-2}_{j=0}(-1)^j
\left( \begin{array}{c}
m \\ j+1
\end{array} \right)
x_{lim,k-j}
={S}^{m-1}_1{G}^{0} (x_{lim,k})
$$ \vspace*{-6pt}
\begin{equation}
+\sum^{k-1}_{j=1}{S}^{m-1}_{j+1}{G}^{0} (x_{lim,k-j})
+\sum^{l-1}_{j=k}{S}^{m-1}_{j+1}{G}^{0} (x_{lim,k-j+l}), \  m-1<k<l,
\label{LimDifferencesPmm1}
\end{equation} 
$$ 
x_{lim,m}+(-1)^{m}x_{lim,l}+
\sum^{m-1}_{j=1}(-1)^j
\left( \begin{array}{c}
m \\ j
\end{array} \right)
x_{lim,m-j}
={S}^{m-1}_1{G}^{0} (x_{lim,m-1})
$$ \vspace*{-6pt} 
\begin{equation} 
+\sum^{m-2}_{j=1}{S}^{m-1}_{j+1}{G}^{0} (x_{lim,m-1-j})
+ \!
\sum^{l-1}_{j=m-1}{S}^{m-1}_{j+1}{G}^{0} (x_{lim,m-1-j+l}), \  k=m-1,
\label{LimDifferencesPmm2}
\end{equation} 
$$ 
x_{lim,k+1}+(-1)^{m}x_{lim,l+k-m+1}-
\sum^{k-1}_{j=1}(-1)^j
\left( \begin{array}{c}
m \\ j+1
\end{array} \right)
x_{k-j}
$$ \vspace*{-6pt} 
$$
-\sum^{m-2}_{j=k}(-1)^j
\left( \begin{array}{c}
m \\ j+1
\end{array} \right)
x_{lim,l+k-j}
={S}^{m-1}_1{G}^{0} (x_{lim,k})
+\sum^{k-1}_{j=1}{S}^{m-1}_{j+1}{G}^{0} (x_{lim,k-j})
$$ \vspace*{-6pt} 
\begin{equation} 
+\sum^{l-1}_{j=k}{S}^{m-1}_{j+1}{G}^{0} (x_{lim,k-j+l}), \   0<k<m-1,
\label{LimDifferencesPmm3}
\end{equation} 
and
\vskip -12pt 
\begin{equation}
\sum^{l}_{j=1}{G}^0(x_{lim,j})=0,
\label{ClosePfm1}
\end{equation}
where $m=\lceil \alpha \rceil$ and $G^0(x)$ is defined by Eq.~(\ref{G}).

Coefficients $S_{i,l}$ for any positive integer $l$ and integer $1 \le i \le l$ are slowly converging sums defined by the following equations 
\vspace*{-12pt} 
{\setlength\arraycolsep{0.5pt}
\begin{eqnarray}
&&{S}^{m-1}_{j+1,l}=\sum^{\infty}_{k=0}\Bigl[
{U}^{m-1}_{\alpha} (lk+j) - {U}^{m-1}_{\alpha} (lk+j+1)\Bigr], \  \ 
\label{SerPm}
\end{eqnarray}
}
\vspace*{-8pt}
$$0\le j<l.$$
They satisfy a property
\vskip -12pt
 \begin{equation}
\sum^{l}_{j=1}{S}^{m-1}_{j,l}=0.
\label{SsumPn}
\end{equation}

$U^{m-1}_{\alpha}$ is defined as
{\setlength\arraycolsep{0.5pt}
\begin{eqnarray}
&&U^{m-1}_{\alpha}(n)=U^{m-2}_{\alpha}(n)-U^{m-2}_{\alpha}(n-1 ) \nonumber \\
&&=\Delta^{m-1} U_{\alpha}(n-m+1).
\label{UtildeM}
\end{eqnarray}
}
We already postulate that
\begin{equation}
U_{\alpha}(n)=0, \  \ {\rm when} \ \ n<1.
\label{UNeg}
\end{equation}

In fractional maps,
{\setlength\arraycolsep{0.5pt}
\begin{eqnarray}
&& U^{m-1}_{\alpha}(n)=
\begin{array}{c}
\left\{
\begin{array}{lll}
\sum^{m-1}_{i=0}(-1)^i
\left( \begin{array}{c}
m-1 \\ i
\end{array} \right)
(n-i)^{\alpha-1} \   \
{\rm for} \ \ n > m-1,
\\
\sum^{n-1}_{i=0}(-1)^i
\left( \begin{array}{c}
m-1 \\ i
\end{array} \right)
(n-i)^{\alpha-1} \   \
{\rm for} \  \ n < m.
\end{array}
\right.
\end{array}
\label{UtildeMFr}
\end{eqnarray}
}

In fractional difference maps,  
{\setlength\arraycolsep{0.5pt}
\begin{eqnarray}
&&U^{m-1}_{\alpha}(n)  = \frac{\Gamma(\alpha) \Gamma(n+\alpha-m)}{\Gamma(\alpha-m+1) \Gamma(n)}.
\label{UtildeMFrDif}
\end{eqnarray}
}
For large $n$, $U^{m-1}_{\alpha}(n)\sim n^{\alpha-m}$ in both fractional and fractional difference cases.

\subsubsection{Calculation of the sums $S^{m-1}_{j,l}$} \label{sec:4.1.2}

Calculation of the sums $S^{m-1}_{j,l}$, in the case of fractional difference maps, is quite simple. As it is shown in \cite{Avigail} (Eq.~(4.32) in that paper),
{\setlength\arraycolsep{0.5pt}
\begin{eqnarray}
&&S^{m-1}_{j+1,l}(\alpha)=(\alpha-1)^{(m-1)}S_{j+1,l}(\alpha-m+1)
, \  \  \  0 \le j<l,
\label{St1vsS1FDm}
\end{eqnarray}
}
where we explicitly show the dependence of the sums on $\alpha$ by adding it as the argument, and $0<\alpha-m+1<1$.

It is shown in \cite{TooSimple} that in fractional difference maps with $0<\alpha<1$
{\setlength\arraycolsep{0.5pt}  
\begin{eqnarray} 
&&S_{p,2n}= S^0_{p,2n}=\frac{\Gamma(\alpha) 2^{-\alpha}}{n}(-1)^p\Bigl[1
\nonumber \\
&&+2\sum^{n-1}_{j=1}(\cos(\pi j/(2n)))^{1-\alpha}\cos(\pi j(2p+\alpha-3)/(2n)) \Bigr], \  \ 0<p \le 2n
\label{Sp_2n}
\end{eqnarray}
}
and
{\setlength\arraycolsep{0.5pt}   
\begin{eqnarray} 
&&S_{p,2n+1}= S^0_{p,2n+1}=\frac{2^{2-\alpha}\Gamma(\alpha)}{2n+1}(-1)^p   
\sum^{n-1}_{j=0}\Bigl(\cos\frac{\pi (2j+1)}{2(2n+1)} \Bigr)^{1-\alpha}
(-1)^j
\nonumber \\
&&\times
\sin\frac{\pi (2j+1)(2p+2n-2+\alpha)}{2(2n+1)}, \  \ 0<p \le 2n+1.
\label{Sp_2n+1} 
\end{eqnarray}
}

The calculations are more complicated in the case of fractional maps. It would take a full page to write the equations which allow fast and accurate numerical computations of sums for arbitrary $l$-cycles of any order $\alpha$ (see Eqs.~(4.33),~(4.44),~and~(4.36) in \cite{Avigail}). 
In the case $m=1$ ($0<\alpha \le 1$), the equations obtained in \cite{Cycles} are 
{\setlength\arraycolsep{0.5pt}   
\begin{eqnarray} 
&&{S}_{1,l}={S}^0_{1,l}=-1 + \sum^{N}_{k=1}\Bigl[
(lk)^{\alpha-1} - (lk+1)^{\alpha-1} \Bigr]
\nonumber \\
&&+(1-\alpha)l^{\alpha-2}\Biggl\{\zeta_N(2-\alpha) +\frac{\alpha-2}{2l}\Biggl[\zeta_N(3-\alpha)
\nonumber \\
&& +\frac{\alpha-3}{3l}\Biggl(\zeta_N(4-\alpha)+\frac{\alpha-4}{4l}\zeta_N(5-\alpha)\Biggr)\Biggr]\Biggr\}
\nonumber \\
&&+O(N^{\alpha-5})
\label{S1nfl}
\end{eqnarray}
}
and for $0<j<l$
{\setlength\arraycolsep{0.5pt}   
\begin{eqnarray} 
&&{S}_{j+1,l}={S}^0_{j+1,l}=\sum^{N}_{k=0}\Bigl[
(lk+j)^{\alpha-1} - (lk+j+1)^ {\alpha-1} \Bigr]
\nonumber \\
&&+(1-\alpha)l^{\alpha-2}\Biggl\{\zeta_N(2-\alpha) +\frac{\alpha-2}{2l}\Biggl[(2j+1)\zeta_N(3-\alpha)
\nonumber \\
&& +\frac{\alpha-3}{3l}\Biggl((3j^2+3j+1)\zeta_N(4-\alpha)+\frac{(\alpha-4)}{4l}(2j+1)
\nonumber \\
&&\times (2j^2+2j+1)\zeta_N(5-\alpha)\Biggr)\Biggr]\Biggr\}
+O(N^{\alpha-5}).
\label{Sjnl}
\end{eqnarray}
}

\subsubsection{Poincar\'{e} plots}
\label{sec:4.1.3}

\begin{figure}[b]
\sidecaption
\includegraphics[scale=.34]{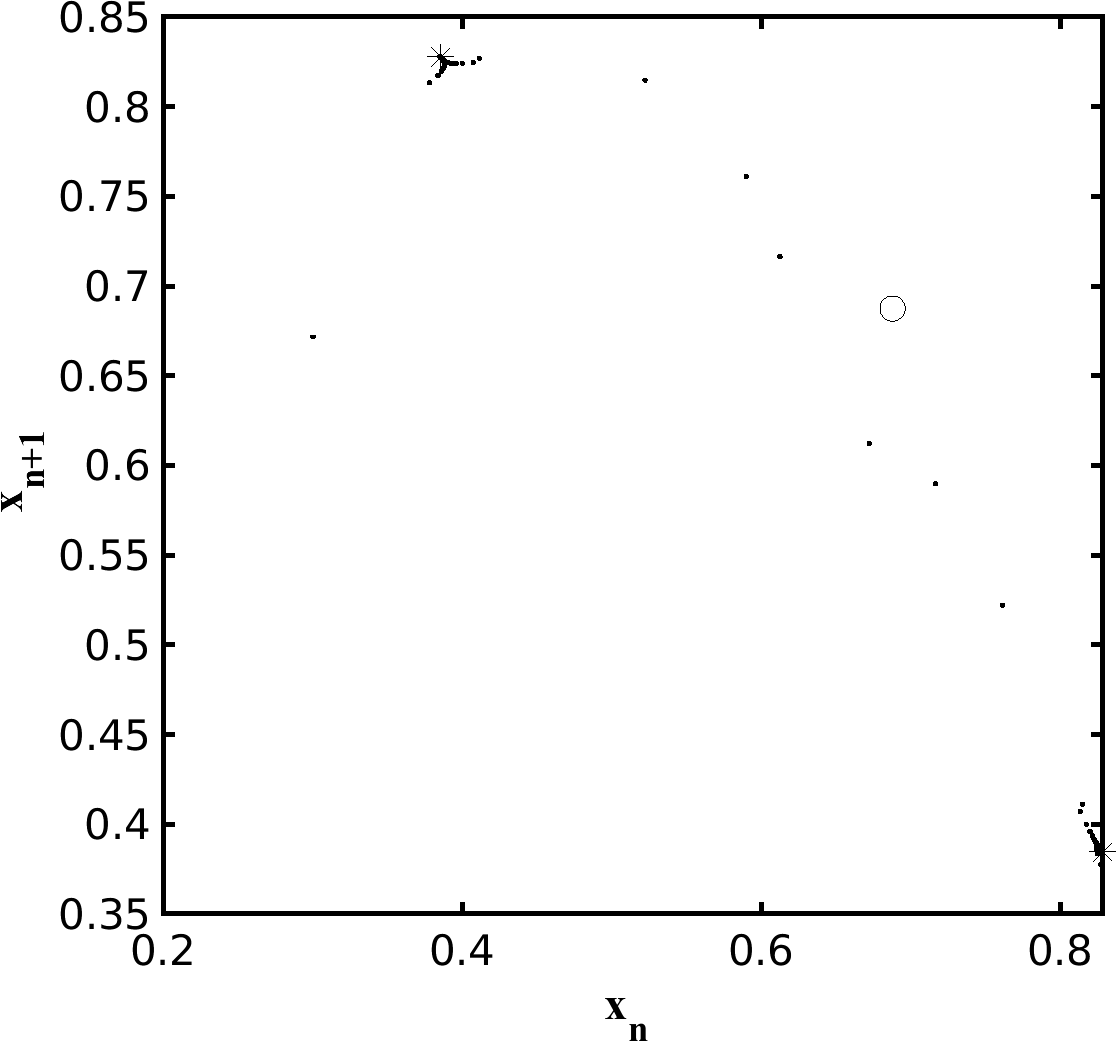}
\vspace{-0.25cm}
\caption{The Poincar\'{e} plot (500000 iterations) for fractional difference logistic map with $\alpha=0.75$, $K=3.2$, $h=1$, and the initial condition $x_0=0.3$. The asymptotically stable $T=2$ sink is marked by the stars and the unstable fixed point $(K-1)/K$ by the circle.
This figure is reprinted from \cite{Cycles} with the permission of Springer Nature.}
\label{fig4}
\end{figure}
\begin{figure}[b]
\sidecaption
\includegraphics[scale=.34]{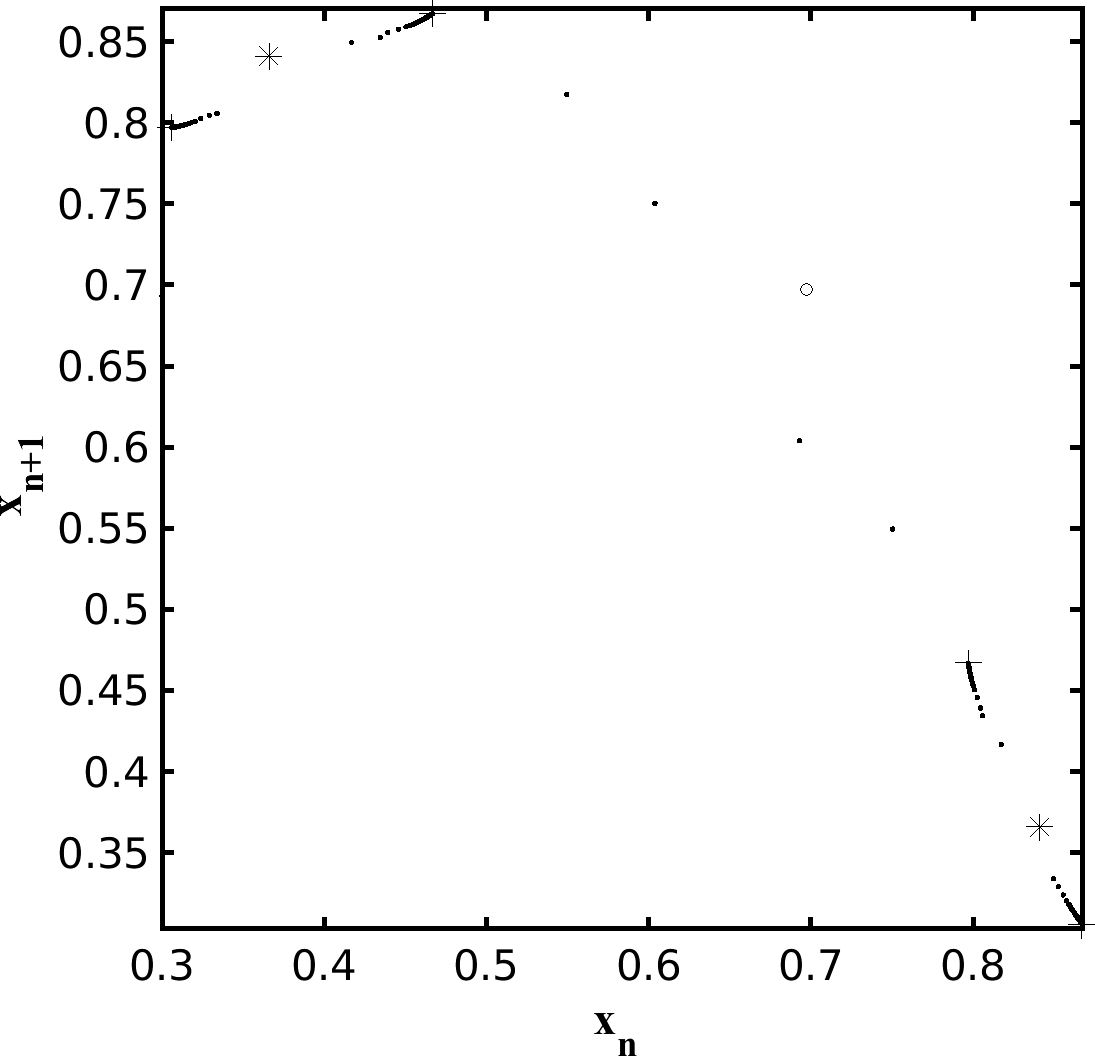}
\vspace{-0.25cm}
\caption{ The Poincar\'{e} plot (500000 iterations) for fractional difference logistic map with $\alpha=0.75$, $K=3.3$, $h=1$, and the initial condition $x_0=0.3$. Stable $T=4$ sink marked by the plus signs. The asymptotically unstable $T=2$ sink is marked by the stars and the unstable fixed point $(K-1)/K$ by the circle.
This figure is reprinted from \cite{Cycles} with the permission of Springer Nature.}
\label{fig5}
\end{figure} 
\begin{figure}[b]
\sidecaption
\includegraphics[scale=.34]{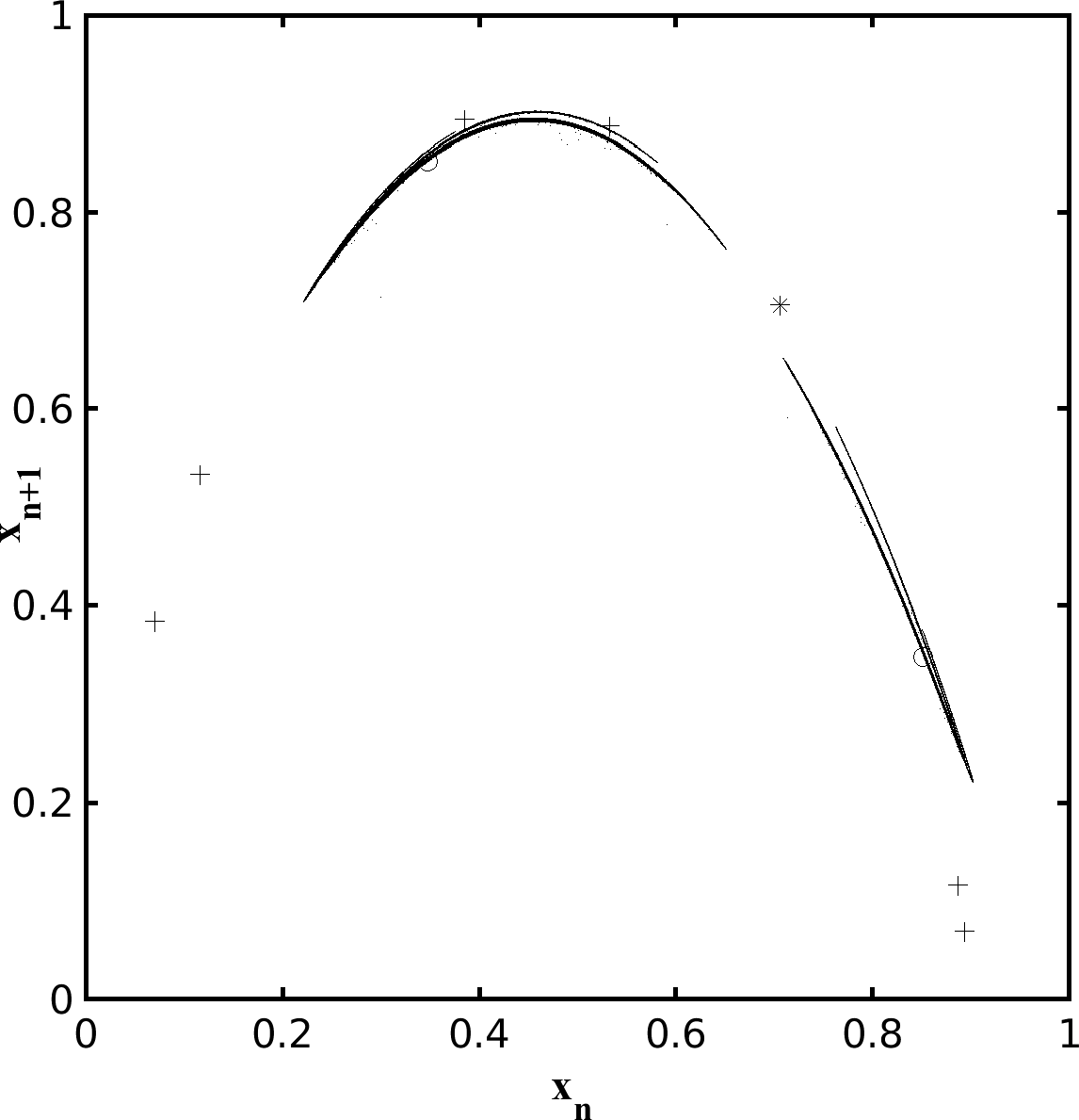}
\vspace{-0.25cm}
\caption{The Poincar\'{e} plot (500000 iterations) for fractional difference logistic map with $\alpha=0.75$, $K=3.4$, $h=1$, and the initial condition $x_0=0.3$. The asymptotically unstable $T=2$ sink, \{0.348,0.852\} is marked by the circles and the unstable fixed point $(K-1)/K=0.706$ by the star. Two asymptotically unstable $T=3$ cycles, \{0.116,0.533,0.887\} and \{0.0696,0.385,0.894\}, are marked by the plus signs. This figure is reprinted from \cite{Cycles} with the permission of Springer Nature.}
\label{fig6}
\end{figure}
\begin{figure}[b]
\sidecaption
\includegraphics[scale=.34]{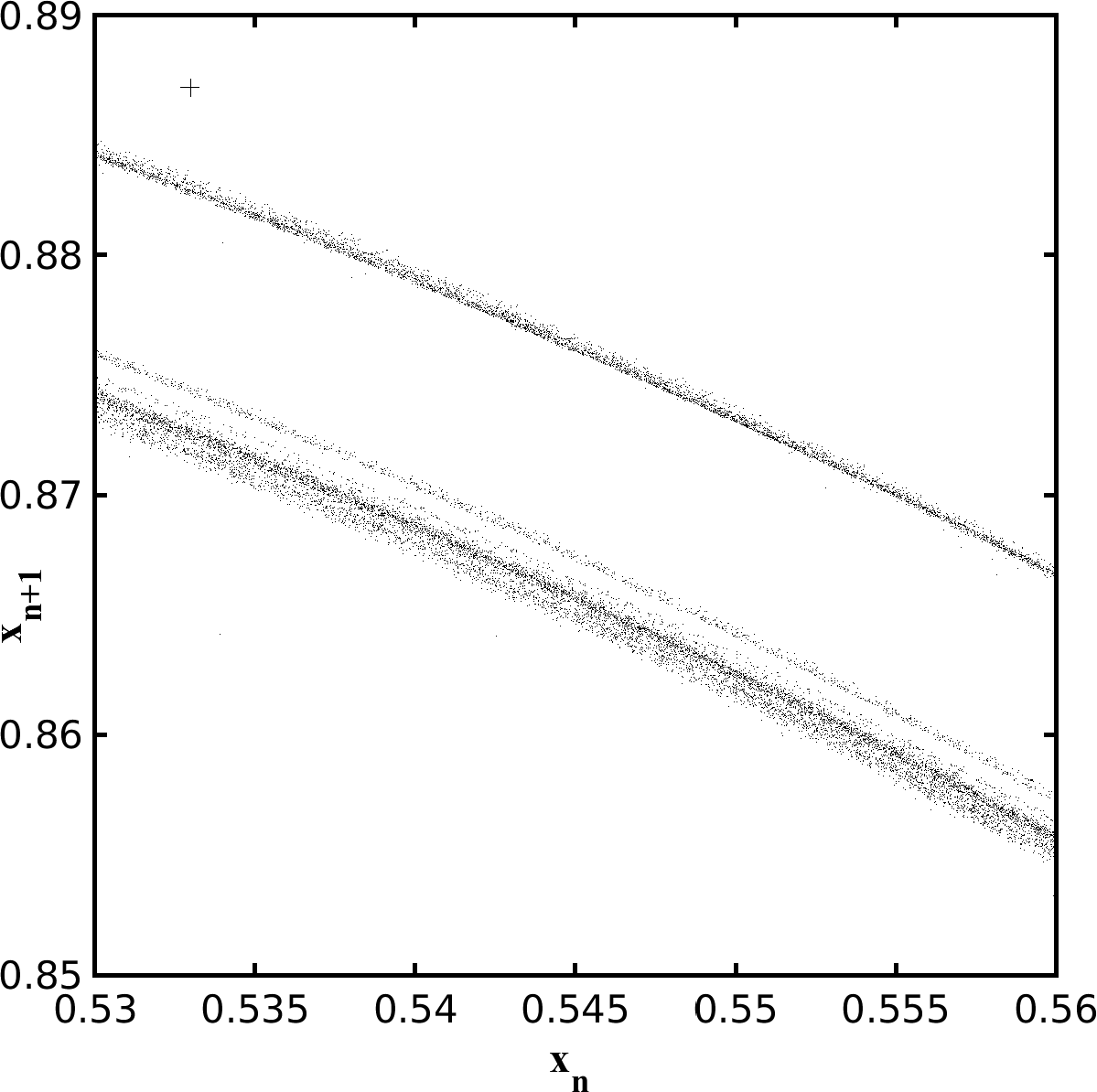}
\vspace{-0.25cm}
\caption{A zoom of Fig.~\ref{fig6} with $\alpha=0.75$, $K=3.4$, $h=1$, and the initial condition $x_0=0.3$. This figure is reprinted from \cite{Cycles} with the permission of Springer Nature.}
\label{fig7}
\end{figure}
In \cite{Cycles}, the asymptotically periodic behavior in the FDLM was analyzed using Poincar\'{e} plots (return maps). Return maps for the FDLM with $\alpha=0.75$, various values of $K$, and the initial point $x_0=0.3$, obtained after 500000 iterations, are given in Figs.~\ref{fig4}--\ref{fig6}. Calculated using Eq.~(\ref{eqT2logSolu}), $T=2$ sink is stable in Fig.~\ref{fig4} and unstable in Figs.~\ref{fig5}~and~\ref{fig6}. It is easy to see that the return map in Fig.~\ref{fig5} converges to the $T=4$ point calculated using Eqs.~(\ref{LimDifferencesPmm1})~and~(\ref{ClosePfm1}). This confirms the correctness of the equations and calculations.

Similar to regular dynamics \cite{Cvitanovic}, asymptotically periodic trajectories should represent 
the skeleton of fractional chaos.   
The formulae defining periodic points can be used to analyze not only stable asymptotically periodic solutions but also chaos in discrete fractional systems.  
The chaotic return map in Fig.~\ref{fig6} is similar to a multi-scroll attractor of a dissipative system.   
The fact that fractional systems behave like dissipative systems
was noticed by many authors. One of the first examples of this similarity can be found in \cite{ZSE}. 
Fig.~\ref{fig7} demonstrates a quasi-fractal structure of the Poincar\'{e} plot in Fig.~\ref{fig6}. A consistent quantitative analysis of chaos in discrete fractional systems is still an open problem.

\subsubsection{Bifurcation points}
\label{sec:4.1.4}

Equations that define asymptotic bifurcation points in any generalized fractional maps, including the GFLM, were derived in \cite{Bif}. They are defined by the following theorem:
\begin{theorem}\label{The3} 
The asymptotic $T=2^{n-1}$ --  $T=2^{n}$ bifurcation points, $2^{n-1}$ values of $x_{2^{n-1}bif,i}$ with $0<i \le 2^{n-1}$ and the value of the nonlinear parameter $K_{2^{n-1}bif}$, of a fractional generalization of a nonlinear one-dimensional map $x_{n+1}=F_K(x_n)$ written as the Volterra difference equations of convolution type
\begin{eqnarray}
x_{n}= x_0 
-\sum^{n-1}_{k=0} G^0(x_k) U_\alpha(n-k),
\label{FrUUMapN1}
\end{eqnarray} 
where $G^0(x)=h^\alpha G_K(x)/\Gamma(\alpha)$, $x_0$ is the initial condition, $h$ is the time step of the map, $\alpha$ is the order of the map, $G_K(x)=x-F_K(x)$, $U_\alpha(n)=0$ for $n \le 0$, $U_\alpha(n) \in \mathbb{D}^0(\mathbb{N}_1)$, and
\begin{eqnarray}
&&\mathbb{D}^i(\mathbb{N}_1) \ \ = \ \ \left\{f: \left|\sum^{\infty}_{k=1}\Delta^if(k)\right|>N, 
\right.\nonumber \\
&& \left.\forall N, \ \ N \in \mathbb{N}, 
\sum^{\infty}_{k=1}\left|\Delta^{i+1}f(k)\right|=C, \ \ C \in \mathbb{R}_+\right\},
\label{DefForm1}
\end{eqnarray}
are defined by the system of $2^{n-1}+1$ equations   
\begin{eqnarray} 
&&x_{2^{n-1}bif,m+1}-x_{2^{n-1}bif,m}=S_{1, 2^{n-1}}    
G^0(x_{2^{n-1}bif,m})\nonumber\\
&&\hspace{25pt}{}+\sum^{m-1}_{j=1}S_{j+1,2^{n-1}} G^0(x_{2^{n-1}bif,m-j})\nonumber \\
&&\hspace{25pt}{}+\sum^{2^{n-1}-1}_{j=m}S_{j+1,2^{n-1}} G^0(x_{2^{n-1}bif,m-j+2^{n-1}}),\label{LimDifferencesNN}\\
&&\hspace{150pt} 0<m<2^{n-1},\nonumber
\\
&&\sum^{2^{n-1}}_{j=1} G^0(x_{2^{n-1}bif,j})=0,
\label{LimDifferencesNN2}
\\
&& \operatorname{det}(A)=0
\label{DetNN},
\end{eqnarray}
where  
\begin{eqnarray} 
&&S_{j+1,l}=\sum^{\infty}_{k=0}\Bigl[
U_{\alpha} (lk+j) - U_{\alpha} (lk+j+1)\Bigr], \nonumber\\
&&\hspace{50pt} 0 \le j<l, \quad
S_{i,l}=S_{i+l,l}, \quad i \in \mathbb{Z},
\label{SerNN1}
\end{eqnarray}
and the elements of the $2^{n-1}$-dimensional matrix $A$ are 
\begin{eqnarray} 
&&\hspace{-10pt}A_{i,j}=\frac{dG^0(x)}{dx}\Bigl|_{x_{2^{n-1}bif,j}}
 \sum^{i+2^{n-1}-1}_{m=i}S_{m-j+1,2^n}+\delta_{i,j}, 
 \label{DetNN2} 
 \\
 &&\hspace{140pt} 0 < i,j \le 2^{n-1}. \nonumber
\end{eqnarray}
\end{theorem}
Theorem~\ref{The3} was used to calculate the bifurcation points in the
FDLM (Table~\ref{table:T1}) and the FLM (Table~\ref{table:T2}). Further analysis of these tables will be presented in Section~\ref{sec:4.5}.
\begin{table*}[ht!]
\centering
\caption{Approaching the Feigenbaum constant $\delta$ in the regular and fractional difference logistic maps. $K_1(n)$ are the values of the map parameter for the period $2^{n-1}$ -- period $2^{n}$ bifurcation points in the regular logistic map ($\alpha=1$); $K_{0.5}(n)$ are the same (asymptotic) points in the case $\alpha=0.5$; $\delta_1=\Delta K_{1}(n-2)/\Delta K_{1}(n-1)= [K_{1}(n-1)-K_1(n-2)]/[K_{1}(n)-K_1(n-1)]$; $\delta_{0.5}$ is the same value in the case $\alpha=0.5$; $\delta_F=\delta=4.6692016$.
This table is reprinted from \cite{Bif} with the permission of AIP Publishing.}
\begin{tabular}{|  c  |  c  |  c  |  c  |  c  |  r  |  r  |}
\hline 
$\ n\ $ & $K_1(n)$ & $K_{0.5}(n)$ & $\ \frac{\Delta K_{1}(n-2)}{\Delta K_{1}(n-1)}\ $ 
     & $\ \frac{\Delta K_{0.5}(n-2)}{\Delta K_{0.5}(n-1)}\ $ & $\delta_1-\delta_F\ \ $ & $\delta_{0.5}-\delta_F\ \ $  \\ \hline
 1 & $3$                    & $2.414213562373$  &          &          &         & \\ \hline
 2 & $\ 3.44948974278317\ $ & $\ 3.031508074451\ $  &          &          &         & \\ \hline
 3 & $3.54409035955192$ & $3.128802940099$   & $\ 4.751446\ $ & $\ 6.344574\ $ & $8.2\cdot 10^{-2}$ & $1.7$  \\ \hline
 4 & $3.56440726609543$ & $3.152254569285$  & $4.656251$ & $4.148747$ & $-1.3\cdot 10^{-2}$  & $-5.2\cdot 10^{-1}$  \\ \hline
 5 & $3.56875941954382$ & $3.156981864934$  & $4.668242$ & $4.960898$ & $-9.6\cdot 10^{-4}$ & $2.9\cdot 10^{-1}$  \\ \hline
 6 & $3.56969160980139$ & $3.158020855572$  & $4.668739$ & $4.549892$ & $-4.6\cdot 10^{-4}$ & $-1.2\cdot 10^{-1}$  \\ \hline
 7 & $3.56989125937812$ & $3.158240983307$  & $4.669132$ & $4.719944$ & $-6.9\cdot10^{-5}$ & $5.1\cdot 10^{-2}$  \\ \hline
 8 & $3.56993401837397$ & $3.158288326235$  & $4.669183$ & $4.649644$ & $-1.9\cdot10^{-5}$ & $-2.0\cdot 10^{-2}$ \\ \hline  
 9 & $3.56994317604840$ & $3.158298449625$  &  $4.669198$ & $4.676588$ &   $-4\cdot10^{-6}$     & $7.4\cdot 10^{-3}$  \\ \hline
10 & $3.56994513734217$ & $3.158300619014$  &    & $4.666470$ &           & $-2.7\cdot 10^{-3}$ \\ \hline
11 &             & $3.158301083532$ &       & $4.670197$ &          & $1.0\cdot 10^{-3}$ \\ \hline
12 &             & $3.158301183025$ &       & $4.668838$ &          & $-3.6\cdot 10^{-4}$ \\ \hline
\end{tabular}
\label{table:T1}
\end{table*}

\begin{table}[ht!]
\centering
    \caption{The order $\alpha=0.5$ FLM:  the values of the map parameter for the asymptotic bifurcation points $K_{0.5}$ and the ratios $\delta_{0.5}=\Delta K_{0.5}(n-2)/\Delta K_{0.5}(n-1)= [K_{0.5}(n-1)-K_{0.5}(n-2)]/[K_{0.5}(n)-K_{0.5}(n-1)]$
converging to the Feigenbaum number $\delta_F=\delta=4.6692016$.
This table is reprinted from \cite{Bif} with the permission of AIP Publishing.}
\begin{tabular}{| c  | c  |  c  | r |}
\hline 
$n$ & $K_{0.5}(n)$  &  $\ \frac{\Delta K_{0.5}(n-2)}{\Delta K_{0.5}(n-1)}\ $ & $\delta_{0.5}-\delta_F\ $  \\ \hline
 1  & $\ 3.930166681941\ $ &          &              \\ \hline
 2  & $4.886148479108$ &          &              \\ \hline
 3  & $5.057125821136$ & $\ 5.591278\ $ & $9.2\cdot 10^{-1}$\\ \hline
 4  & $5.096340664045$ & $4.360016$ & $-3.1\cdot 10^{-1}$\\ \hline
 5  & $5.104461683175$ & $4.828808$ & $1.6\cdot 10^{-1}$\\ \hline
 6  & $5.106226043488$ & $4.602812$ & $-6.6\cdot 10^{-2}$\\ \hline
 7  & $5.106601705434$ & $4.696670$ & $2.7\cdot 10^{-2}$\\ \hline
 8  & $5.106682343101$ & $4.658641$ & $-1.1\cdot 10^{-2}$\\ \hline
 9  & $5.106699598575$ & $4.673164$ & $4.0\cdot 10^{-3}$\\ \hline
10  & $5.106703295325$ & $4.667742$ & $-1.5\cdot 10^{-3}$\\ \hline
   \end{tabular}
\label{table:T2}
\end{table}

\subsection{Stability}
\label{sec:4.2}

Generalized fractional maps are Volterra difference equations of convolution type with power-law-like kernels. Stability and existence of periodic solutions in Volterra difference equations of convolution type with various kernels were investigated even before the introduction of fractional maps (see Section 6.3 of the textbook \cite{Ela} and monographs \cite{Gil,Kost}). The results in these books and in paper \cite{Minh} were presented in the abstract mathematical form.
The problem of the linear asymptotic stability of the fixed points of the discrete convolution equations with arbitrary kernels was investigated in \cite{Ela1} (for more publications on the topic see 
\cite{Ela,Ela2,Ela3,Ela4,StabO}). Semi-analytic and numerical results for the case of the fractional (power-law kernel) standard and logistic maps were obtained in \cite{ME2,ME5} (see also reviews \cite{ME6,HBV4}).
The problem of linear asymptotic stability of fractional difference equations (falling factorial kernels) of the orders $0<\alpha<1$ was investigated later (see, \cite{St1,St2,St3}), and, more recently, the case of complex order fractional difference equations was investigated in \cite{St4}. It was found that the fixed points in the complex fractional order FDLM are stable when the map parameter $K$ (for the $x=0$ fixed point) or $2-K$ (for the $(K-1)/K$ fixed point) is inside of a boundary curve which parametric representation is
 {\setlength\arraycolsep{0.5pt}   
\begin{eqnarray} 
&&\gamma(t)= \Bigl(Re\Bigl[2^\alpha\Bigl(\sin\frac{t}{2}\Bigr)^\alpha
e^{il[\frac{\alpha\pi}{2}+t(1-\frac{\alpha}{2})]} \Bigr]+1,
\nonumber \\
&&Im\Bigl[2^\alpha\Bigl(\sin\frac{t}{2}\Bigr)^\alpha
e^{i[\frac{\alpha\pi}{2}+t(1-\frac{\alpha}{2})]} \Bigr] \Bigr), \  \ t\in[0,2\pi].
\label{CO} 
\end{eqnarray}
}

In \cite{ME15}, the conditions of stability of fixed points in generalized fractional maps of the orders $0<\alpha<1$ Eq.~(\ref{FrUUMapN1}) were formulated as the following theorem: 
\begin{theorem}\label{T4} 
The map Eq.~(\ref{FrUUMapN1}), where $G^0(x)=h^\alpha G_K(x)/\Gamma(\alpha)$, $x_0$ is the initial condition, $h$ is the time step of the map, $\alpha$ is the order of the map, $G_K(x)$ is a nonlinear function depending on the parameter $K$, $U_\alpha(n)=0$ for $n \le 0$, and $\Delta U_\alpha(n) \in l_1$ is asymptotically stable if and only if the conditions 
\begin{eqnarray}
&&0<\frac{d G(x)}{dx}\Big|_{x=x_f}<\frac{\Gamma(\alpha)}{S_{2,2} h^\alpha},
\label{ZeqNlambdaStabilityN}
\end{eqnarray} 
where $S_{2,2}=\sum^{\infty}_{k=1}U_\alpha(k)(-1)^{k+1}$,  are satisfied.
\end{theorem}

The GFLM has two fixed points $x_{f1}=0$ and $x_{f2}=(K-1)/K$.
In the case of the GFLM, the conditions of stability may be written as 
\begin{eqnarray}
&&1-\frac{\Gamma(\alpha)}{S_{2,2} h^\alpha}<K<1
\label{LogStab0}
\end{eqnarray}
for the $x_{f1}$ fixed point and 
\begin{eqnarray}
&&1<K<\frac{\Gamma(\alpha)}{S_{2,2} h^\alpha}+1
\label{LogStabX2}
\end{eqnarray}
for $x_{f2}$.
The bifurcation point is 
\begin{eqnarray}
&&K_c=\frac{\Gamma(\alpha)}{S_{2,2} h^\alpha}+1.
\label{LogFirstBif}
\end{eqnarray}   

In the case of the FDLM, when $S_{2,2}=\Gamma(\alpha)2^{-\alpha}$, the stability conditions are:
\begin{eqnarray}
&&1-\Bigl(\frac{2}{h}\Bigr)^\alpha <K<1
\label{LogStab0N}
\end{eqnarray}
for the $x_{f1}$ fixed point and 
\begin{eqnarray}
&&1<K<\Bigl(\frac{2}{h}\Bigr)^\alpha +1,
\label{LogStabX2N}
\end{eqnarray}
for $x_{f2}$.
The bifurcation point is 
\begin{eqnarray}
&&K_c=\Bigl(\frac{2}{h}\Bigr)^\alpha+1.
\label{LogFirstBifn}
\end{eqnarray}

In \cite{PerD3}, the authors derived the conditions of asymptotic stability for the asymptotically $T=2$ points in fractional difference maps with $h=1$ of the orders $0<\alpha<1$. They showed that in the FDLM, the asymptotically $T=2$ point is asymptotically stable if and only if
\begin{eqnarray}
&& 1+2^\alpha<K<1+\sqrt{2^\alpha+2^{1+2\alpha}-2^{1+3\alpha/2}\sin(\alpha\pi/4)}.
\label{T2Bah}
\end{eqnarray}

\subsection{Cascade of bifurcations type trajectories (CBTT)}
\label{sec:4.3}

\begin{figure}[!t]
\includegraphics[width=0.9\textwidth]{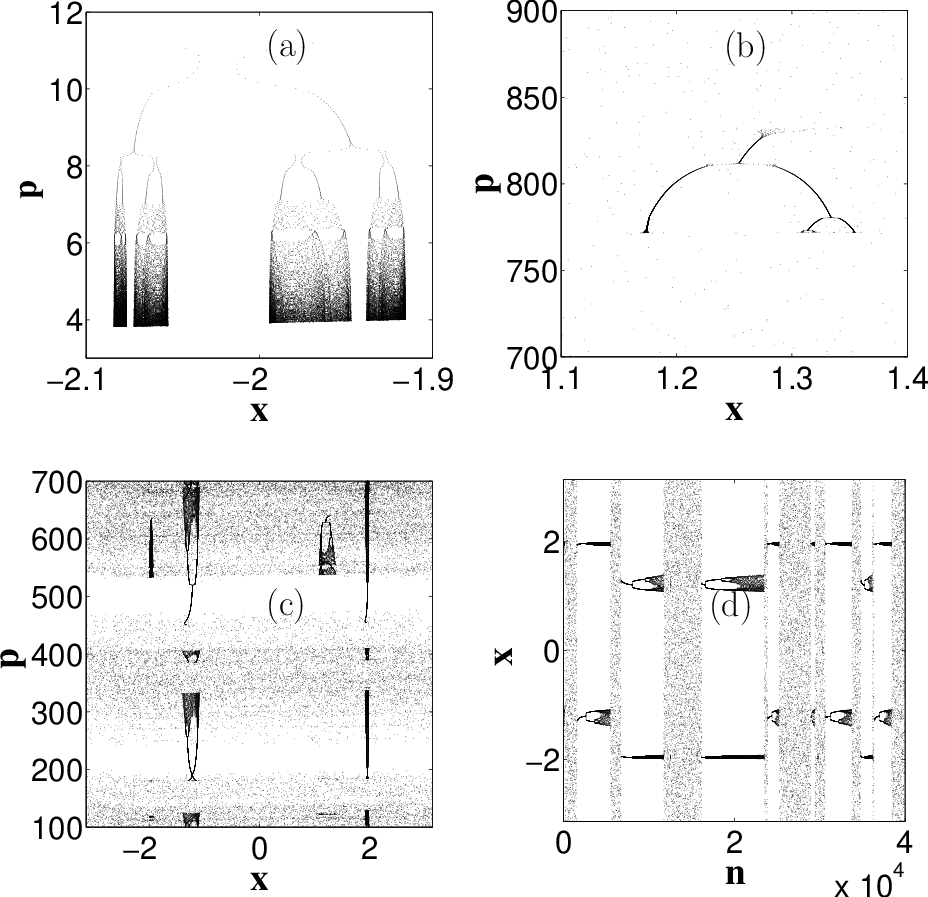}
\vspace{-0.25cm}
\caption{A single CBTT in the RLFSM Eqs.~(\ref{2})~and~(\ref{3}). (a) One of the two branches of
the CBTT for  $\alpha=1.1, K=3.5$. 
(b) A zoom of a small feature in an intermittent 
trajectory for  $\alpha=1.95, K=6.2$.
(c) An intermittent trajectory in phase space for 
$\alpha=1.65, K=4.5$. 
(d) $x$ of $n$ for the case (c).   This figure is reprinted from \cite{ME5} with the permission of AIP Publishing.
}
\label{fig8}
\end{figure}
\begin{figure}[!t]
\includegraphics[width=0.9\textwidth]{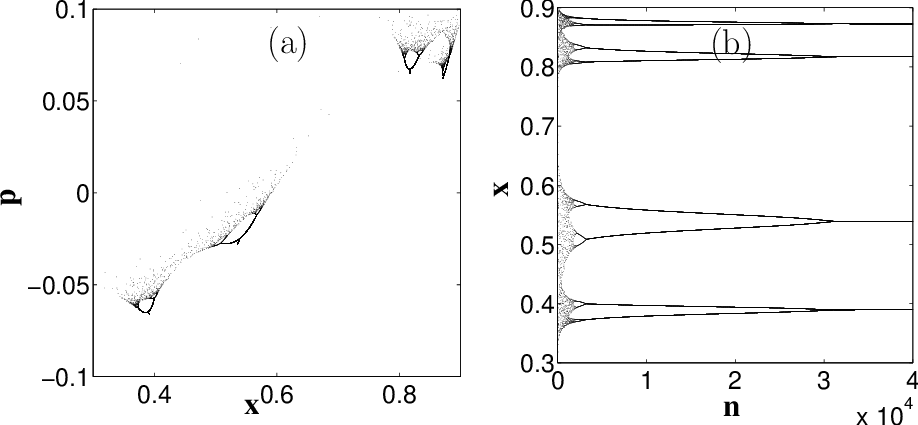}
\vspace{-0.25cm}
\caption{An inverse CBTT in the Caputo LFM Eqs.(\ref{LMCx})~and~(\ref{LMCp}) with $\alpha=1.2$, $K=3.45$.
40000 iterations on a trajectory with $x_0=0.01$ and $p_0=0.1$. (a) Phase
space. (b) $x-n$ graph. This figure is reprinted from \cite{ME5} with the permission of AIP Publishing.
}
\label{fig9}
\end{figure}
One of the most interesting and ubiquitous features of fractional and fractional difference maps is cascade of bifurcations type trajectories (CBTT). They were first noticed in the first paper in which the Riemann-Liouville fractional standard map (RLFSM), where $G_K(x)=K\sin(x)$, was investigated \cite{ME2} (see Fig.~5 in that paper). 
The FRLSM is defined as
\begin{equation}
\label{2}
p_{n+1} = p_n - K \sin x_n ,
\end{equation}
\begin{equation}
x_{n+1} = \frac{1}{\Gamma (\alpha )} 
\sum_{i=0}^{n} p_{i+1}
[(n-i+1)^{\alpha -1}-(n-i)^{\alpha -1}]
, \ \ \ \ ({\rm mod} \ 2\pi ).
\label{3}
\end{equation}
In CBTTs, cascades of bifurcations occur on single trajectories. A trajectory may start converging to a low periodicity trajectory but then bifurcates (the number of bifurcations may be from one to infinity) and end converging to a higher periodicity or chaotic trajectory. In inverse CBTTs, the evolution is opposite: a trajectory may start as a chaotic or converging to a high periodicity trajectory but then, through a series of mergers, ends converging to a low periodicity trajectory. The examples of CBTTs in the RLFSM and in the Caputo FLM of the orders $1<\alpha<2$ are presented in Fig.~\ref{fig8} and Fig.~\ref{fig9}.

CBTTs in the FLM and the FDLM of the orders $0<\alpha<1$ were investigated in papers \cite{ME5} (Fig.~3f there), \cite{ME5n} (Fig.~8d there), \cite{Chaos2018} (Figs.~1~and~2 there), review \cite{HBV4}, and \cite{Die} (Figs.~6.2,~6.6,~6.7~and~6.9 there). Two examples of CBTTs in the Caputo FLM of the order $\alpha=0.1$ are presented in Fig.~10.
\begin{figure}[!t]
\begin{center}
\includegraphics[width=0.9\textwidth]{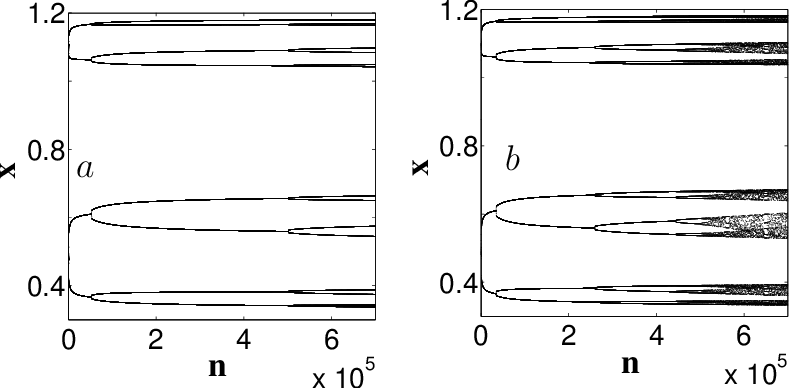}
\vspace{-0.25cm}
\caption{ 
Cascade of bifurcations type trajectories (CBTT) in the
Caputo FLM Eq.(\ref{FrCMapLogLow}) with $h=1$. In both cases $\alpha=0.1$ and $x_0=0.001$: 
(a) in the case $K=22.37$, the last bifurcation from the period  
$T=8$ to the period $T=16$ occurs after approximately $5 \times 10^5$ iterations; 
(b) in the case $K=22.423$, after approximately 
$5 \times 10^5$ iterations the trajectory becomes chaotic. This figure is reprinted from \cite{Chaos2018}, 
with the permission of AIP Publishing.
}
\end{center}
\label{fig10}
\end{figure}
\begin{figure}[!t]
\begin{center}
\includegraphics[width=0.9\textwidth]{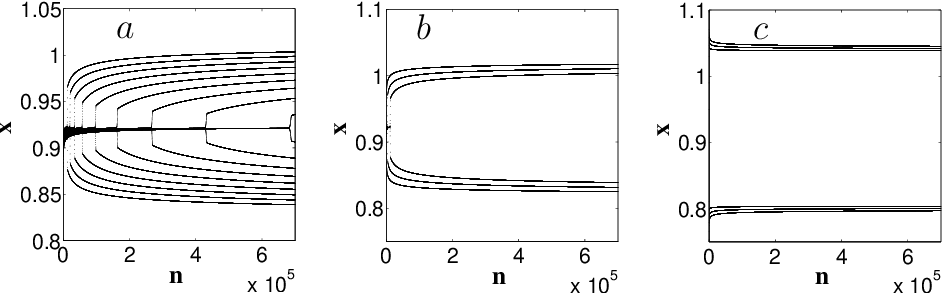}
\end{center}
\caption{Asymptotically period two trajectories for the Caputo FLM Eq.(\ref{FrCMapLogLow})
with $h=1$, $\alpha=0.1$, and $K=15.5$: (a) nine
 trajectories with the initial conditions $x_0=0.29+0.04i$, $i=0,1,...,8$  
($i=0$ corresponds to the rightmost bifurcation);
(b) $x_0=0.61+0.06i$, $i=1,2,3$; (c) $x_0=0.95+0.04i$, $i=1,2,3$. As $ n
\rightarrow \infty $ all trajectories converge to the limiting values
$x_{lim1}=0.80629$ and  $x_{lim2}=1.036030$. 
The unstable fixed point is $x_{lim0}=(K-1)/K=0.93548$.
This figure is reprinted from \cite{Chaos2018}, 
with the permission of AIP Publishing.
}
\label{fig11}
\end{figure}
\begin{figure}[htb]
\centering
  \begin{tabular}{@{}cccc@{}}
    \includegraphics[width=.32\textwidth]{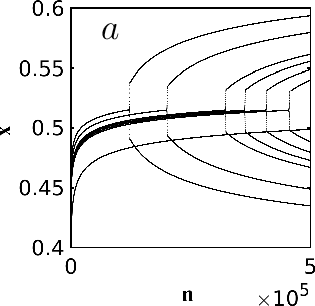} &
    \includegraphics[width=.31\textwidth]{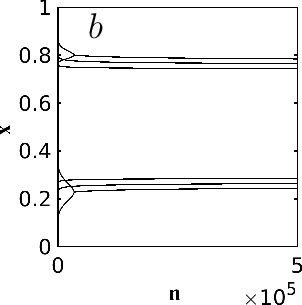} &
    \includegraphics[width=.31\textwidth]{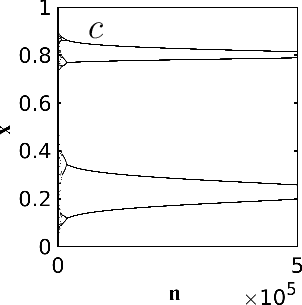} &   
  \end{tabular}
  \caption{Asymptotically period two trajectories for the Caputo FDLM  Eqs.~(\ref{LMlt1}) 
with $h=1$, $\alpha=0.1$, and $K=2.5$: (a) seven trajectories with the initial 
conditions $x_0=0.0001$, $x_0=0.085+0.005i$, $i=0,1,2,3$, $x_0=0.12$, and 
$x_0=0.14$ (the leftmost bifurcation);   
(b) $x_0=0.6+0.1i$, $i=1,2,3$; (c) $x_0=0.9$. As $ n\rightarrow \infty $ 
all trajectories converge to the limiting values
$x_{lim1}=0.3045$ and  $x_{lim2}=0.7242$. 
The unstable fixed point is $x_{lim0}=(K-1)/K=0.6$. 
This figure is reprinted from \cite{Die} with the permission of Springer
Nature.}
\label{fig12}
\end{figure}
\begin{figure}[htb]
\centering
  \begin{tabular}{@{}cccc@{}}
    \includegraphics[width=.49\textwidth]{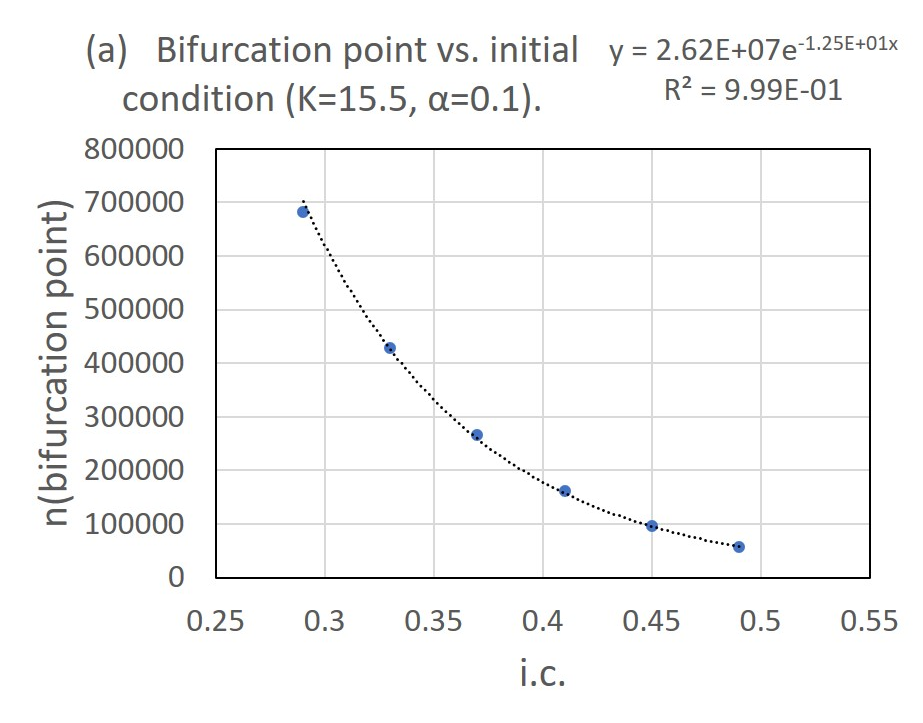} &
    \includegraphics[width=.49\textwidth]{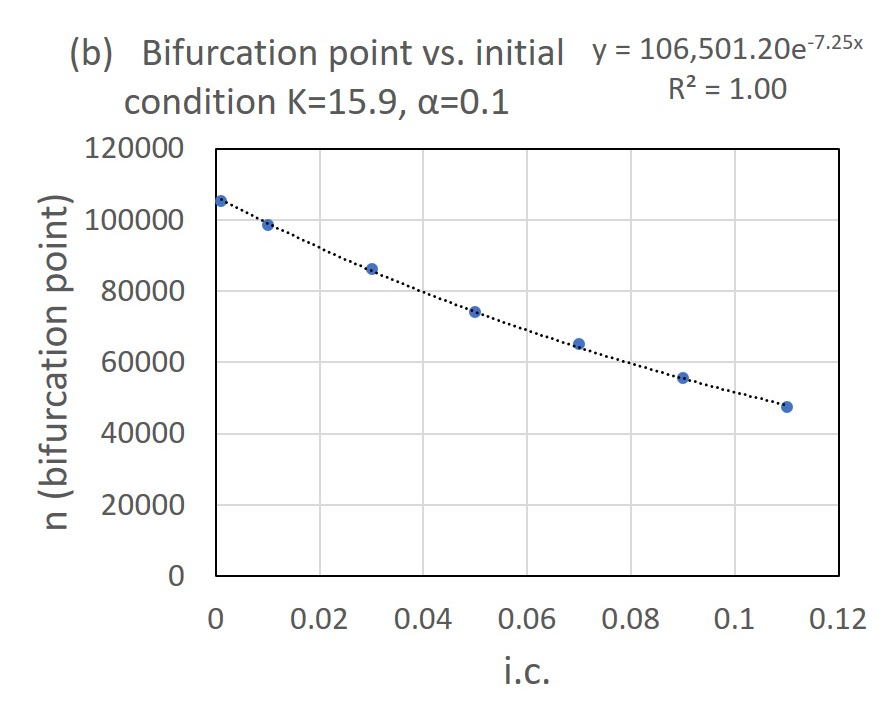} & \\ 
    \includegraphics[width=.49\textwidth]{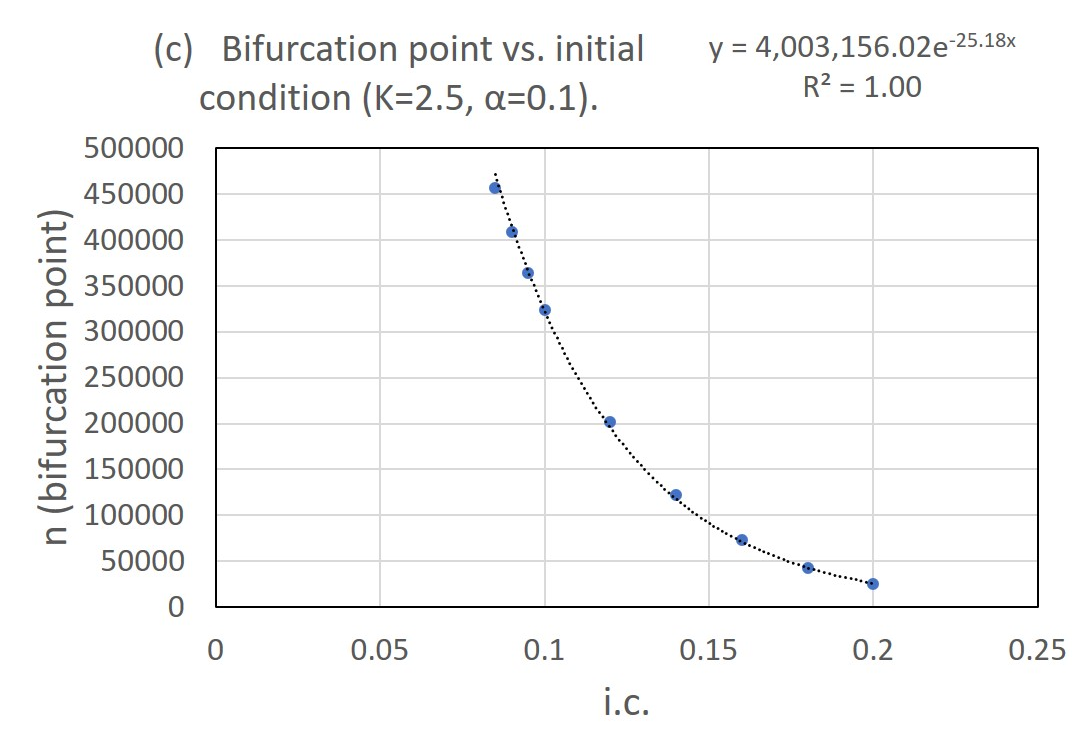} &
    \includegraphics[width=.49\textwidth]{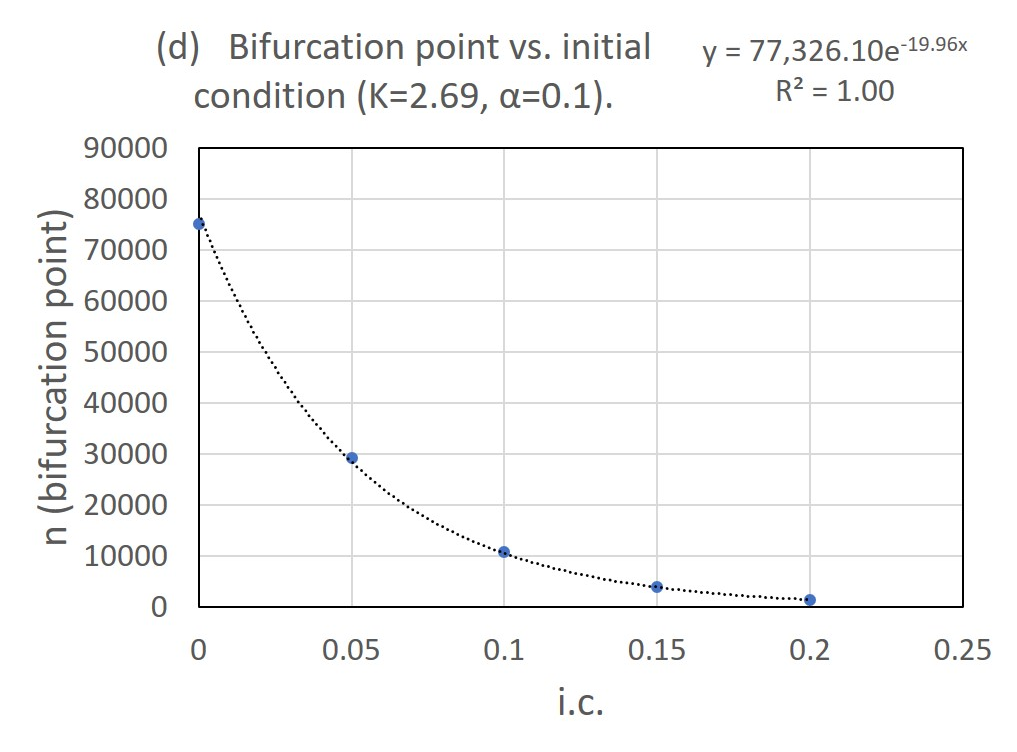} &
  \end{tabular}
  \caption{The time of the regular convergence to a stable fixed point until the bifurcation on a single trajectory for the Caputo FLM (a) and (b) and the Caputo FDLM (c) and (d) as a function of the initial condition. Maps' parameters are indicated on the figures. 
This figure is reprinted from \cite{Die} with the permission of Springer
Nature.}
\label{fig13}
\end{figure}

Whether a periodic trajectory will become a CBTT or an inverse CBTT depends on the initial conditions. 
Figs.~\ref{fig11}~and~\ref{fig12} show this dependence for the Caputo FLM and FDLM.  In the FDLM, when the initial conditions are close to one, the trajectory becomes an inverse CBTT (Fig.~\ref{fig12}c).
As one may see from Figs.~\ref{fig11}a~and~\ref{fig12}a, for small values of the initial conditions, the maps first slowly, in a regular way, start converging to the unstable fixed point but then suddenly become unstable and bifurcate. 
The time (number of iterations) $t$ of the convergence to the unstable fixed point prior to a bifurcation decreases with the increase in the initial conditions. The dependance of this time on the initial conditions, as it is shown in Fig~\ref{fig13}, is exponential, i.e.
\begin{equation}
t=t_me^{-\gamma x},
\label{ExpT}
\end{equation}
where $x$ is the initial condition and constants
$t_m$, which is the maximal possible number of iterations before the bifurcation, and $\gamma$ depend on the map's parameters.

\subsection{Power-law convergence of trajectories}
\label{sec:4.4}

Power-law convergence of trajectories to asymptotically periodic points is another feature that distinguishes fractional maps from regular maps. 
The FLM and the FDLM are particular forms of the generalized fractional maps with asymptotically power-law kernel, which are Volterra difference
equations of convolution type.
It has been proven that if the zero solution of a linear Volterra difference equation of convolution type is uniformly asymptotically stable, then it is exponentially stable if and only if the kernel
decays exponentially (see Theorem 5 in \cite{Ela2}). 

The power-law convergence (which for $0<\alpha<1$ is $n^{-\alpha}$, where $n$ is the number of iterations and $\alpha$ is the fractional order of a map) and divergence of trajectories was semi-analytically and numerically   
analyzed in \cite{ME2} (Section 1 and Fig. 1), \cite{ME3} (see Fig. 1), \cite{ME4} (Section 3.2), 
\cite{ME6} (Section 3.3.3 and figures in it), and \cite{ME7} (see Fig. 5 for $0< \alpha<1$)  for the fractional and fractional difference standard map. 
Power-law convergence in the fractional logistic map was shown in 
\cite{ME5} (see Fig.~6), and for the fractional and fractional difference cases with $0< \alpha<1$, the $n^{-\alpha}$--convergence of trajectories was demonstrated in \cite{Die} (see Fig. 6.8).

A proof that in Caputo fractional difference maps convergence to asymptotically stable fixed points is $O(n^{-\alpha})$ may be found in \cite{Cermak2015}.  
In \cite{Anh}, the authors strictly proved that convergence obeys the power law $\Delta x \sim n^{-\alpha}$, where $0<\alpha<1$ is the order of a fractional difference map. 

This implies that the Lyapunov exponents in converging to the fixed point Caputo fractional difference maps are equal to zero. This means that 
in fractional difference maps the Lyapunov exponent analysis of the convergence to fixed points is unnecessary; it still can be applied in some cases to analyze the divergence of trajectories (see, e.g., \cite{WBLya,DD,GG,JGB}).

\subsubsection{The FLM with $1 < \alpha<2$}
\label{sec:4.4.1}

The power-law convergence of fractional maps' trajectories to asymptotically periodic points was first demonstrated for the fractional standard maps of the orders $1<\alpha<2$ \cite{ME2,ME3,ME4}. Rates of convergence of trajectories to the fixed and periodic points of the Caputo (Eqs.~(\ref{LMCx})~and~(\ref{LMCp})) and Riemann-Liouville (Eqs.~(\ref{FLMRLp})~and~(\ref{FLMRLx})) fractional logistic maps of the orders $1<\alpha<2$ were first investigated in \cite{ME5}. 
\begin{figure}[!t]
\includegraphics[width=0.9\textwidth]{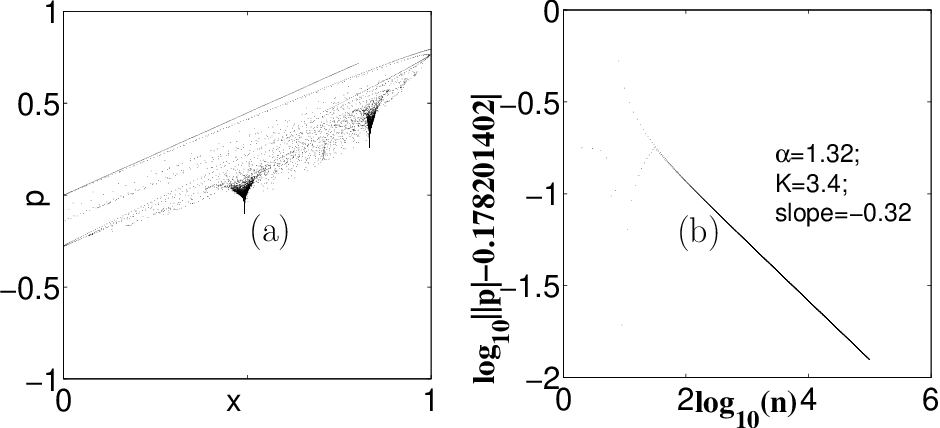}
\vspace{-0.25cm}
\caption{The RLFLM (Eqs.~(\ref{FLMRLp})~and~(\ref{FLMRLx})) with $\alpha=1.32$, $K=3.4$.
(a) Phase space: 300 trajectories with $x_0=0$, $p_0=10^{-6}+0.00024i$, 
$0 \le i <300$. All converging trajectories converge to the $T=2$
antisymmetric in $p$ sink. 
(b) $\log n-\log p$ graph showing the rate of convergence $\delta p
\sim n^{-\alpha +1}$ on a single trajectory. 
This figure is reprinted from \cite{ME5}, 
with the permission of AIP Publishing.
}
\label{fig14}
\end{figure}
In \cite{ME5} the Riemann-Liouville FLM (RLFLM) is defined as
{\setlength\arraycolsep{0.5pt}
\begin{eqnarray}
\label{FLMRLp}
&&p_{n+1} = p_n + K x_n (1-x_n)-x_n ,  \\
&&x_{n+1} = \frac{1}{\Gamma (\alpha )} 
\sum_{i=0}^{n} p_{i+1}[(n-i+1)^{\alpha-1}-(n-i) ^{\alpha-1}].
\label{FLMRLx}
\end{eqnarray}
}
As in the case of the fractional standard maps, the partition of the phase space into
the areas (basins) of convergence to asymptotically stable periodic sinks originating from the 
period one sink $(0,0)$ is almost the same (numerical result) 
for the Caputo FLM and the RLFLM. For  $0<K<1$ 
all converging trajectories converge to $(0,0)$ point as 
$x \sim n^{-\alpha -1}$, $p \sim n^{-\alpha}$. For $1<K<K_{c1L}$ (see Eq.~(\ref{KcL}))
the only stable sink is the period one $((K-1)/K,0)$ sink and the rate of
convergence is $\delta x \sim n^{-\alpha}$, $p \sim n^{-\alpha+1}$.
For   $K_{c1L}<K<K_{c2L}$, where $K_{c2L}$ is the $T=2$ -- $T=4$ asymptotic bifurcation point, all converging  trajectories 
(this is a result from the large number of numerical simulations)
converge, with the same rate (see Fig.~\ref{fig14}b), to the $T=2$
sink antisymmetric in $p$ Fig.~\ref{fig14}a. It is shown in \cite{ME5}, that asymptotically, for large $n$, momentum $p$ obeys the following law:
 \begin{equation} \label{P2Asy} 
p_n=p_l(-1)^n+ \frac{A}{n^{\alpha -1}},
\end{equation}
where
\begin{equation} \label{A} 
A= \frac{K-1+\frac{2\Gamma(\alpha)}{V_{\alpha l}}}{2K\Gamma(2-\alpha)}.
\end{equation}
We have to note that power-law convergence to asymptotically periodic points is much better investigated in the fractional standard map (see \cite{ME2,ME3,ME4,ME5}) than in the fractional logistic map.

\subsubsection{The FLM with $0 < \alpha<1$}
\label{sec:4.4.2}

\begin{figure}[htb]
\centering
  \begin{tabular}{@{}cccc@{}}
    \includegraphics[width=.31\textwidth]{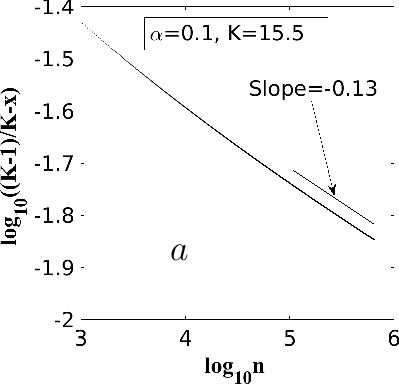} &
    \includegraphics[width=.31\textwidth]{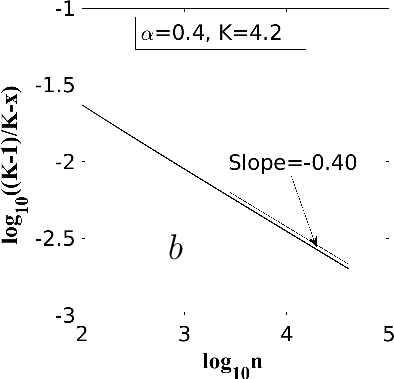} & 
    \includegraphics[width=.31\textwidth]{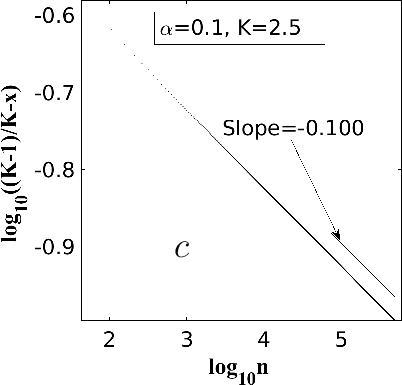} & 
  \end{tabular}
  \caption{Power-law convergence of trajectories to the fixed point (stable (b) or unstable (a) and (c)) for the fractional (a) and (b) and fractional difference (c) Caputo logistic $\alpha$-family of maps. Maps' parameters are indicated on the figures. This figure is reprinted from \cite{Die} with the permission of Springer Nature.
}
\label{fig15}
\end{figure}

As it was shown in \cite{Anh}, convergence to fixed points in the FDLM follows the power law $n^{-\alpha}$. For the FLM, the same law of convergence was numerically shown in \cite{Die} (see Figs.~6.8a and Figs.~6.8b in that paper). It is interesting that in CBTTs initial convergence to unstable fixed points also follows the same law.
This also was demonstrated in Figs.~6.8a and Figs.~6.8c from \cite{Die}. Figs.~6.8 from \cite{Die} is reprinted here as Fig.~\ref{fig15}.

\subsection{Asymptotic cascades of bifurcations and Feigenbaum constant $\delta$}
\label{sec:4.5}

Various forms of fractional maps were investigated and used in applications in hundreds of papers. According to the Web of Science, only one paper \cite{FallLog} with the title \textit{Discrete fractional logistic map and its Chaos} was cited  444 times. In most of those papers, the authors, based on the calculations on single trajectories, constructed bifurcation diagrams to demonstrate maps' complex dynamics and the cascade of bifurcation scenario of transition to chaos. The described in previous sections features of fractional and fractional difference maps make this construction complicated, and it became a subject of multiple discussions. Besides, it should be emphasized that, as mentioned in Section~\ref{sec:4}, fractional and fractional difference maps have no periodic points except fixed points and all bifurcation diagrams are asymptotic bifurcation diagrams. 

\subsubsection{Bifurcations diagrams based on the iterations on single trajectories}
\label{sec:4.5.1}

As we already mentioned in Section~\ref{sec:3.1}, when bifurcation diagrams are based on iterations on single trajectories, it is important to eliminate all transient processes. Because convergence of trajectories to asymptotically periodic points in fractional and fractional difference maps follows the power law $n^{-\alpha}$, it requires many thousands of iterations to obtain reasonable approximations of asymptotic bifurcation diagrams. It is simply impossible to obtain good approximations for maps of small orders. If the maps' order is 0.1, then it would require $10^{10}$ iterations to reduce the initial deviation from a periodic point by a factor of ten. To the best of our knowledge, the largest number of iterations used in simulations on single trajectories of fractional or fractional difference maps was of the order $10^{6}$. The longer computer runs are impossible due to the computer memory restrictions. The dependence of the quality of bifurcation diagrams obtained after $10^4$ iterations on single trajectories on $\alpha$ for the FLM may be easily seen if one compares the diagram for $\alpha=0.5$ in Fig.~3.b from the 2013 paper \cite{ME5} to the diagram for $\alpha=0.1$ in Fig.~3.d from that paper (the initial condition is $x_0=0.1$ in both cases). 

\begin{figure}[!t]
\includegraphics[width=1\textwidth]{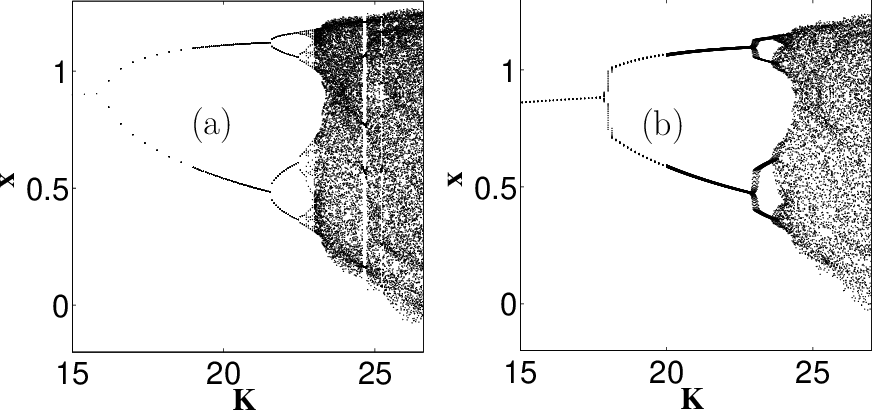}
\vspace{-0.25cm}
\caption{Dependence of bifurcation diagrams of the fractional maps 
on the number of iterations on a single trajectory used in their calculation.  
Bifurcation diagrams for the FLM with $\alpha=0.1$ and $x_0=0.1$.
(a) 10000 iterations on each trajectory. 
(b) 100 iterations on each trajectory. This figure is reprinted from \cite{ME6} with the permission of Springer
Nature.
}
\label{fig16}
\end{figure}

In Fig.~8 from the January 2013 paper \cite{ME5n}, the author compared the FLM with $\alpha=0.1$ and $x_0=0.1$ bifurcation diagrams obtained after $10^4$ iterations (Fig.~8g in that paper) to the diagram obtained after 100 iterations (Fig.~8h). One may easily see that an increase in the number of iterations leads to a huge shift of the bifurcation diagram to the left. This shift may be easily explained by the fact that the trajectories with $x_0=0.1$ are CBTTs and those that after 100 iterations converge to the fixed point may later bifurcate and start converging to the asymptotically $T=2$ trajectory. The corresponding figure, Fig.~\ref{fig16}, is reprinted in this publication. 
A strong dependence of bifurcation diagrams on the initial conditions in fractional difference maps was demonstrated in Fig.~6 from \cite{ME7} for the Caputo fractional difference standard map. 
\begin{figure}[!t]
\centering
\includegraphics[width=4in]{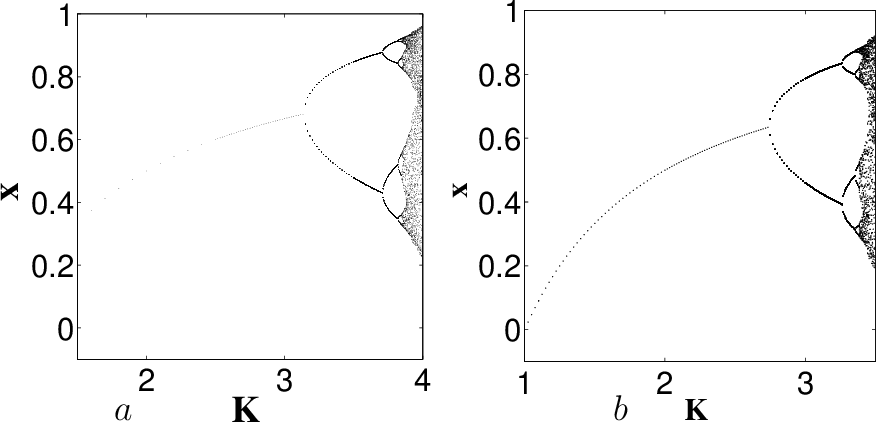}
\caption{  Bifurcation diagrams  for the Caputo LFM ($a$)
 and the Caputo FDLM ($b$) 
with $\alpha=0.8$ obtained after 1000 iterations with the initial
condition $x_0=0.01$. This figure is reprinted from \cite{ME8} with the permission of L\&H Scientific Publishing.
}
\label{fig17}
\end{figure}
In the FLM and the FDLM, the trajectories with the initial conditions close to zero are CBTTs and the corresponding bifurcation diagrams will shift to the left when the number of iterations increases. But the trajectories with the initial conditions close to one are inverse CBTTs and the corresponding bifurcation diagrams when the number of iterations increases, will shift to the right.    

It should be also noted that the FLMs' bifurcation diagrams, if compared to the regular logistic map's diagrams, are stretched along the parameter (horizontal) axis, and the FDLMs' bifurcation diagrams are contracted. The stretchiness and the contraction increase as the order of maps evolves from $\alpha=1$ (regular logistic map) to zero.  This has a simple explanation. The zero order Caputo FLM is identically zero, $x_n=0$, which zero fixed point is unconditionally stable, and the first bifurcation point moves from $K=3$ to $K=\infty$ when $\alpha$ evolves from one to zero. This also may be seen from the condition of stability of the FLM's fixed point Eq.~({\ref{LogStabX2}). Near zero, the gamma function tends to the infinity while the value of $S_{2,2}$ remains finite (see Table~3 from \cite{Cycles}). 
On the other hand, the fixed point -- $T=2$ bifurcation point for the FDLM, which is defined by Eq.~(\ref{LogFirstBifn}), evolves from  $K=3$ to $K=2$ when $\alpha$ evolves from one to zero.


\subsubsection{The exact bifurcation diagrams}
\label{sec:4.5.2}

\begin{figure*}[!t]
\begin{center}                                
\includegraphics[width=0.99 \textwidth]{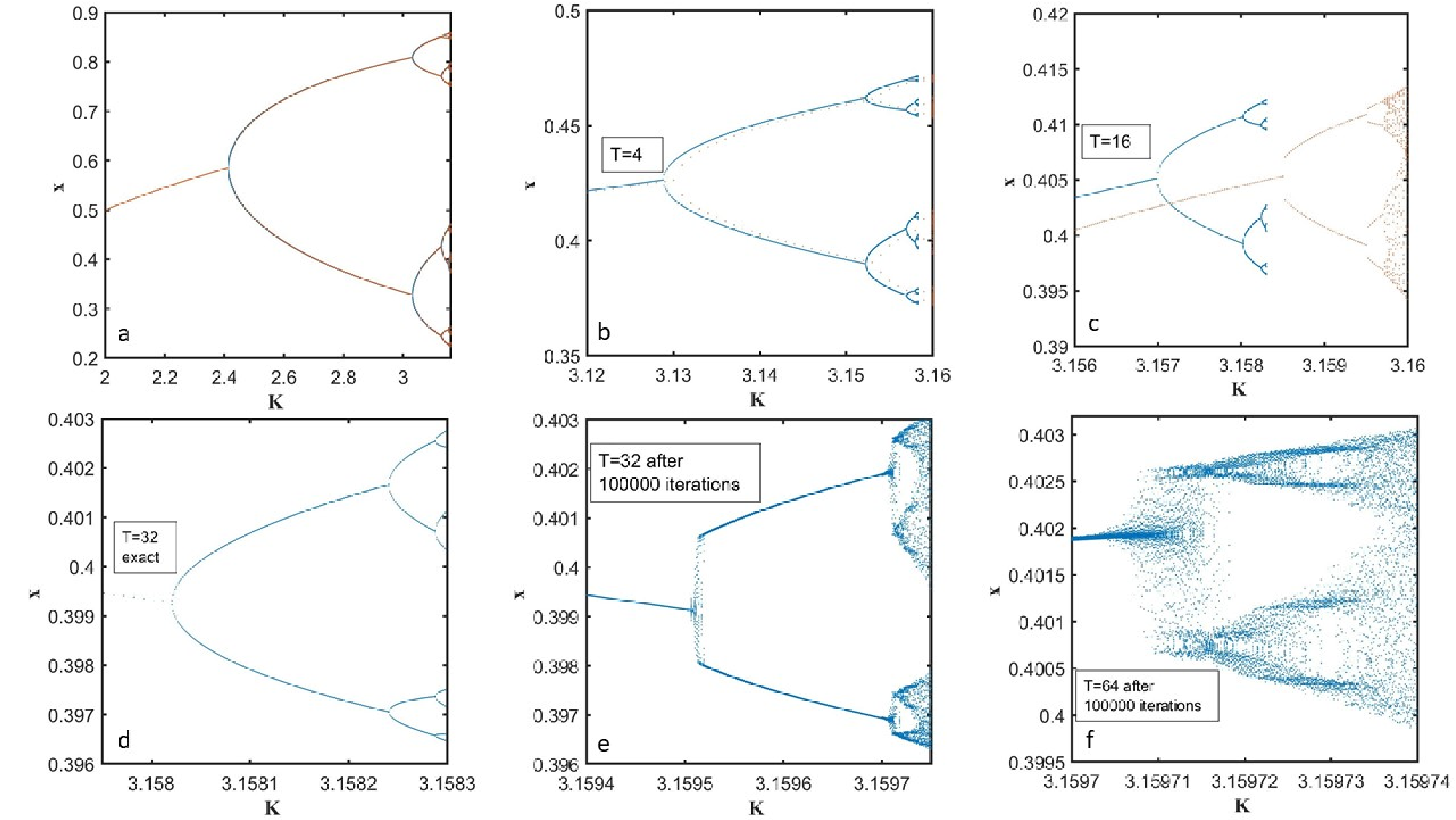}
\end{center}                                
\vspace{-0.25cm}
\caption{A part of the bifurcation diagram for the Caputo fractional
  difference logistic map of the order $\alpha=0.5$ from~$K=2$ (fixed
  point) to approximately~$K=3.16$ ($T=128$ periodic point). In figures
  a--c the steady line represents the solution of  Eqs.~(\ref{LimDifferencesPmm1})-(\ref{ClosePfm1})   
 (the exact solution) and the dots represent numerical calculations on a single trajectory with~$x_0=0.3$ after~$10^5$ iterations. 
Figure~d represents the exact solution from~$T=32$ on the left to~$T=128$
on the right. Figures~e and~f represent the calculations on a single
trajectory.
This figure is reprinted from \cite{Bif}, with the permission of AIP Publishing.
}
\label{fig18}
\end{figure*}
The reprinted here from \cite{Bif} Fig.~\ref{fig18}, the order $\alpha=0.5$ FDLM bifurcation diagrams, were  obtained by solving Eqs.~(\ref{LimDifferencesPmm1})-(\ref{ClosePfm1}) and compared to the bifurcation diagrams obtained after $10^5$ iterations on single trajectories. 

From Fig.~\ref{fig18}b, one may see a noticeable difference between the exact solutions and the results obtained after $10^5$ iterations on a single trajectory already for the period four ($T = 4$) points. For $T > 16$ (Figs.~\ref{fig18}b,~\ref{fig18}e,~and~\ref{fig18}f), the results obtained after $10^5$ iterations on a single trajectories seem to be noticeably inaccurate. Multiple papers investigating various fractional difference maps contain bifurcation diagrams obtained by iterations on a single trajectories, and the number of iterations in all these papers is much less than $10^5$. We must note, that a numerical solution of the exact equations defining periodic and bifurcation points, when the period $T=2^N$ is large, requires solving of $2^N$ nonlinear algebraic equations, and it is not practical to use these equations to construct bifurcation diagrams when $N>10$.

\subsubsection{The Feigenbaum constant $\delta$}
\label{sec:4.5.3}

As it was mentioned at the beginning of Section~\ref{sec:2}, the logistic map was used to investigate the universality in regular dynamics. One of the characteristics of this universality is Feigenbaum constat $\delta$ which describes self-similarity of the intervals of parameters between consecutive bifurcations. Tables~\ref{table:T1}~and~\ref{table:T2}, constructed for the FDLM and the FLM, contain the values of the map parameter $K_1(n)$ (for the regular logistic map) and $K_{0.5}(n)$ (for the order $\alpha=0.5$ FDLM and FLM) at the period $2^{n-1}$ -- period $2^{n}$ bifurcation points.

The results for $n=1,2,\ldots,10$ and~$\alpha=0.5$ were obtained using Eqs.~(\ref{LimDifferencesPmm1})-(\ref{ClosePfm1}). 
The tables also show the corresponding values of the ratios $\delta_1=\Delta K_1(n-2)/\Delta K_1(n-1)= [K_1(n-1)-K_1(n-2)]/[K_1(n)-K_1(n-1)]$ and~$\delta_{0.5}=\Delta K_{0.5}(n-2)/\Delta K_{0.5}(n-1)= [K_{0.5}(n-1)-K_{0.5}(n-2)]/[K_{0.5}(n)-K_{0.5}(n-1)]$. All results for $\alpha=0.5$ in the tables were later calculated using Theorem \ref{The3} (Eqs.~\eqref{LimDifferencesNN}--\eqref{DetNN}), and for the first ten values of $n$, the results are identical. Numerical calculations were performed using MATLAB and the obtained results had higher accuracy than the corresponding solutions of Eqs.~(\ref{LimDifferencesPmm1})-(\ref{ClosePfm1}) obtained using Mathematica.
The values for the regular logistic map are available from many sources (see, e.g., \cite{RegCDB1,RegCDB2}).

The values of $\delta_{0.5}$, as in the case of the regular logistic map, oscillate around the Feigenbaum number. They converge significantly slower and this is expected because, in many cases, the convergence in fractional maps follows the power law while the convergence in regular maps is exponential. From the authors' of \cite{Bif} point of view, the data present the sufficient evidence to make the conjecture that the Feigenbaum number exists in fractional and fractional difference maps and has the same value as in regular maps.

\subsubsection{2D ($\alpha$--$K$) and $\alpha$--$x$   bifurcation diagrams}
\label{sec:4.5.4}

\begin{figure}[!t]
\begin{center}
\includegraphics[width=0.9\textwidth]{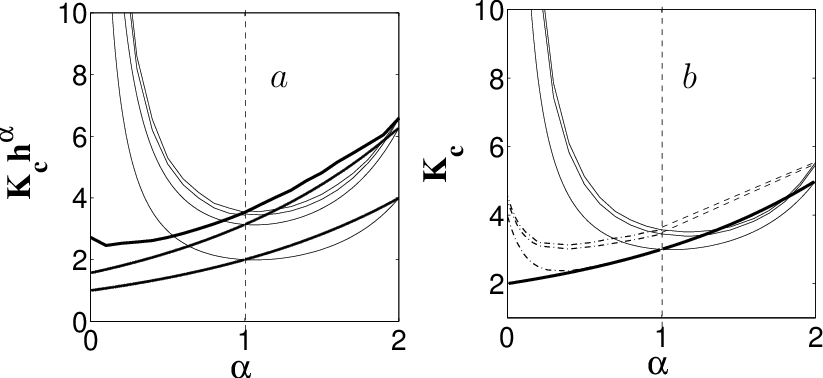}
\vspace{-0.25cm}
\caption{  
2D bifurcation diagrams for the fractional (solid thing lines) and
fractional difference (bold and dashed lines) Caputo standard (a) 
and $h=1$ logistic (b) maps. The first bifurcation, transition from a stable fixed point to a stable period two ($T=2$) sink, 
occurs on the bottom curves.
A $T=2$ sink (in the case of the standard $\alpha$-families of maps
an antisymmetric T=2 sink with $x_{n+1}=-x_n$)
is stable between the bottom and the middle curves. Transition 
to chaos occurs on the top curves. For the standard fractional map transition from a $T=2$ to $T=4$ sink
occurs on the line below the top line (the third from the bottom line). 
Period doubling bifurcations leading to chaos occur in the narrow band between the two top curves.
This figure is reprinted from \cite{Chaos2018} with the permission of AIP Publishing.   
}
\end{center}
\label{fig19}
\end{figure}

As we already mentioned, regular, $K$--$x$, bifurcation diagrams depend on $\alpha$. Bifurcation diagrams for various values of $\alpha$ may be combined in 2D bifurcation diagrams presented here in Fig.~(\ref{fig19}) reprinted from \cite{Chaos2018} for the fractional and fractional difference logistic and standard maps. We should note that these bifurcation diagrams, except for the fixed point--$T=2$ bifurcation points, were obtained by simulations on single trajectories and their accuracy near $\alpha=0$ is low. Nevertheless, these 2D bifurcation diagrams give a qualitatively correct description of the asymptotic dynamical behavior of the logistic and standard maps.
\begin{figure}[!t]
\centering
\includegraphics[width=3.5in]{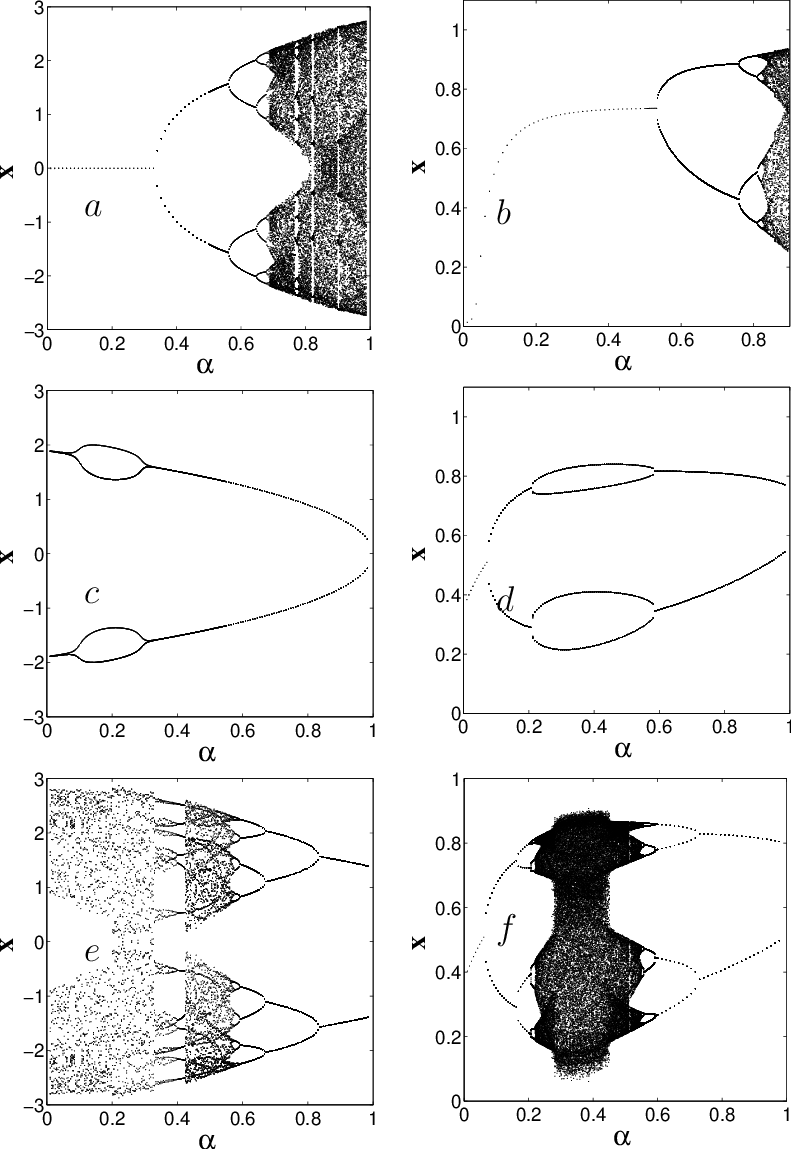}
\vspace{-0.25cm}
\caption{The memory $(\alpha,x)$-bifurcation diagrams for fractional Caputo 
standard (a) and logistic (b) maps and for fractional difference Caputo standard
(c) and (e) and logistic (d) and (f) maps obtained after 5000 iterations.
$K=4.2$ in (a), $K=3.8$ in (b), $K=2.0$ in (c), $K=3.1$ in (d), $K=2.8$ in (e),
and   $K=3.2$ in (f).  This figure is reprinted from \cite{MEBr} 
with the permission of Springer Nature.  
}
\label{fig20}
\end{figure}

It is easy to note by looking at the 2D bifurcation diagrams, that systems with power-law memory demonstrate bifurcations with changes in the memory parameter $\alpha$ when the nonlinearity parameter $K$ stays constant. This property of systems with power-law memory is demonstrated in Fig.~\ref{fig20}. Live species are systems with power-law memory and the dependance of their behavior on memory may be used in biological applications. It, for example, may be related to the explanation of how changes/failures in live biological species can be caused by changes in their memory and nervous system. This also may be used to explain how some diseases may be treated by treating the nervous system.  

\section{Applications of the FLM and the FDLM}
\label{sec:5}

Memory is an important property of many natural and socioeconomic systems. Systems with memory are also widely used in engineering. In many cases, this memory is asymptotically power-law memory. Maps with memory are used to describe discrete systems with memory, to model continuous systems with memory, and all numerical schemes to solve fractional differential equations may be considered as maps with memory. 

A review of the economic applications of maps with memory may be found in \cite{T1E}, and paper \cite{T3E} is dedicated exclusively to the economic applications of the FLM. 
There are many papers in which fractional maps are used to model memristors (see, e.g, \cite{memr}), and in \cite{memrL}, the authors used the logistic map in their discrete fracmemristor model.
Another application of fractional maps is to encrypt images and signals (see, e.g. \cite{crypto1}), and in \cite{crypto2}, the authors used fractional logistic map to do the encryption.

Fractional calculus is frequently used in biological applications in which memory plays a significant role. The following review shows the importance of nonlinear fractional (with power-law memory) dynamics described by fractional differential/difference equations of the orders $0<\alpha<2$, especially when $\alpha$ is close to zero, in biological application.

Memory, as a significant property of humans, is a subject of extensive biophysical and psychological research. The power-law forgetting, the decay of the accuracy on memory tasks as $\sim t^{-\beta}$, with $0<\beta<1$, has been demonstrated in experiments described in  
\cite{Kahana,Rubin,Wixted1,Wixted2,Adaptation1}. 
Human learning is also characterized by power-law memory: the reduction in reaction times that comes with
practice is a power function of the number of training trials \cite{Anderson}.
Dynamics of biological systems at levels ranging from single ion channels up to human psychophysics in \cite{Adaptation1,Adaptation3,Adaptation4,Adaptation2,Adaptation5, Adaptation6} is described by the application of power-law adaptation.

The reasons for human's power-law memory are related to the power-law memory of its building blocks, from individual neurons and proteins to tissues of individual organs. 
It has been shown in \cite{Neuron3,Neuron4,Neuron5}, that processing of the external stimuli by individual neurons, can be described by fractional differentiation with the orders of derivatives $\alpha \in [0,1]$. E.g., in the case of neocortical pyramidal neurons, this order is $\alpha \approx 0.15$.
The power-law memory kernel with the exponent $-0.51 \pm 0.07$ is demonstrated in fluctuations within single protein molecules (see \cite{Protein}).
 
Viscoelasticity is one of the most important applications of fractional calculus. Viscoelastic materials act as substances with power-law memory and their behavior can be described by the fractional differential equations. Viscoelasticity of human tissues was demonstrated in many publications. It is related to the viscoelasticity of the components of tissues: structural proteins, cells, extracellular matrices, and so on. 
References related to various organs may be found in \cite{ME6}. For more recent results on viscoelasticity of the brain tissue see \cite{TissueBrain3}, cardiovascular tissues \cite{CardVasc}, human tracheal tissues \cite{Trac}, human skin \cite{Skin}, human bladder tumours \cite{BTum}, human vocal fold tissues \cite{Voc}, and many other publications.

Applications of fractional calculus in biology also include fractional wave-propagation in biological tissues \cite{TissueWaves1,TissueWaves2,TissueWaves3,TissueWaves4}, 
bioengineering (bioelecrodes, biomechanics, and bioimaging) \cite{Magin}, 
population biology, and epidemiology \cite{PopBioBook2001,HoppBook1975}).

The persistence of power-law memory in biological systems allowed the author of \cite{Die} to describe living species by introducing a  
fitness function, understood as the total (potential) number of descendants produced by a certain age, which evolution can be described by the FLM or the FDLM. Because all living species are asymptotically unstable, the value of the nonlinear parameter was selected from the region of instability of the fixed point. The initial value of fitness was selected near zero. It has been shown that for all values of parameters and various perturbations, lifespans of living species are consistent with the Gompertz distribution and a limited lifespan. This result is consistent with the recent findings \cite{LS} that the human lifespan is really limited. Statistical analysis sets this limit at approximately 114 years. Does this mean that, whatever happens, human life cannot exceed 
this limit? Apparently not. According to the Guinness World Records and the Gerontology Research Group, the longest living person whose dates of birth and death were verified (although not without controversies), Jeanne Calment (1875--1997), lived to 122. We also may note that in the recent paper \cite{FDLMcontrol}, the authors showed that the unstable fixed point of the FDLM may be controlled. This means that if the model introduced in \cite{Die} is correct, there still is a chance to extend human life. 

And I wish Albert Luo to live happily and be healthy to 120.

\section{Conclusion}
\label{sec:6}

In this review, the author tried to outline what is known about the fractional generalizations of the logistic map. Various versions of these generalizations, like the FLM and the FDLM, demonstrate some common finite time general properties which are more complicated than the properties of the regular logistic map. They include persistence of CBTTs, power-law convergence to asymptotically periodic point, and dependance of bifurcation diagrams on the initial conditions and the number of iterations used. 

But the asymptotic properties of the FLM and the FDLM are quite similar to the corresponding properties of the regular logistic map: the periodic and bifurcation points are well defined and do not depend on the initial conditions, and the universality is characterized by the same Feigenbaum constant $\delta$. 

Complicated finite time behavior of the FDLM is used in image and signal inscription applications and for secure communications. It may be used for descriptions of complex economic and biological systems. It is also used in engineering and control.    

Finally, the results of the application of the FDLM to the analysis of human longevity suggest that human life may be prolonged. I hope to live long enough to see the theoretical proof of the universality of fractional dynamics and more applications of the FLM and the FDLM.



%
\begin{acknowledgement}
The author acknowledges continuing support from Yeshiva University and expresses his gratitude to the administration of Courant Institute of Mathematical Sciences at NYU for the opportunity to perform computations at Courant.
\end{acknowledgement}
\ethics{Competing Interests}{The author declares no competing interests.
 }

%
%
%

\end{document}